\documentclass[a4paper,11pt]{article}
\pdfoutput=1 

\usepackage[utf8]{inputenc}
\usepackage[T1,tone,safe]{tipa}
\usepackage{xcolor}
\usepackage{xspace}
\usepackage{lmodern}
\usepackage{latexsym,amsmath,array,amssymb,stmaryrd,theorem,epsfig}
\usepackage{bbm}
\usepackage{pifont}
\usepackage{array}
\usepackage{mathtools}
\usepackage{nicefrac}
\usepackage{slashed}
\usepackage{braket}
\allowdisplaybreaks
\usepackage[utf8]{inputenc}
\usepackage{jheppub}
\usepackage{subdepth}

\newcommand{\tr}{\text{tr}}
\newcommand{\R}{\mathbb{R}}

\newcommand{\Jfrak}{\mathfrak{J}}

\newcommand{\arccot}{\operatorname{arccot}}
\newcommand{\csch}{\operatorname{csch}}
\def \ha {{\frac{1}{2}}}
\def \N {{\mathcal N}}
\def \diag {\operatorname{diag}}

\def \pp {{\rm p}}

\def \hp {{\tfrac{\rm p}{2}}}

\def \LL {{\mathcal{L}}}
\def \DD {{D}}
\def \DDT {{\tilde{\DD}}}

\def \cX {{\cal X}}

\def \cT {{\cal T}}

\def \cY {{\cal Y}}
\def \cS {{\cal S}}

\def \1   {\mathbf{1}}

\def \OO {{\mathcal{O}}}

\def \cD {{\mathcal{D}}}
\def \Id {{\mathbbm{1}}}

\def \d {\textrm{d}}
\def \p {\partial}
\def \pbar{{\bar{\partial}}}
\def \zbar{{\bar{z}}}
\def \g {\gamma}
\def \G {\Gamma}
\def \Gh {\hat{\Gamma}}
\def \eps {\epsilon}
\def \hh {{\rm h}}

\def \td {\tilde}
\def \qt {\td{q}}
\def \sech{\mathrm{sech}}
\def \vp {\varphi}
\def \vt {\vartheta}
\def \th {\theta}
\def \w{\omega}
\def \wt{\td{\w}}
\def \c2{\cos^2\!\vp}
\def \s2{\sin^2\!\vp}
\def \fc2{\frac{1}{\c2}}
\def \fs2{\frac{1}{\s2}}
\def \sfc2{\tfrac{1}{\c2}}
\def \sfs2{\tfrac{1}{\s2}}
\newcommand{\AdS}{\textup{AdS}\xspace}
\newcommand{\CFT}{\textup{CFT}}
\newcommand{\Sphere}{\textup{S}}
\newcommand{\Torus}{\textup{T}}

\def \AdsST  {$\AdS_3 \times \Sphere^3 \times \Torus^4$\xspace}
\def \AdsSSS{$\AdS_3 \times \Sphere^3 \times \Sphere^3 \times \Sphere^1$\xspace}
\newcommand{\alg}[1]{\mathfrak{#1}}
\newcommand{\grp}[1]{\mathrm{#1}}

\newcommand{\su}{\alg{su}}
\newcommand{\SU}{\grp{SU}}

\newcommand{\psu}{\alg{psu}}

\newcommand{\ce}[1]{#1_{\text{c.e.}}}

\newcommand{\fixedspaceL}[2]{\mathrlap{#2}\phantom{#1}}
\newcommand{\fixedspaceR}[2]{\phantom{#1}\mathllap{#2}}		
\def\Zbar{{\bar{Z}}}
\def\hap{\textstyle{p \over 2}}

\def\hap1{\textstyle{p_1 \over 2}}
\def\hap2{\textstyle{p_2 \over 2}}
\def\p{\partial}
\def\pbar{{\bar{\partial}}}
\def\zbar{{\bar{z}}}
\def\lam12{{\lambda_{12} }}
\def\lamP{{\lambda_{P} }}

\title{\boldmath Semiclassical quantization of the mixed-flux $\AdS_3$ giant magnon}
\author{Adam Varga}
\affiliation{Department of Mathematics,\\School of Mathematics, Computer Science \& Engineering,\\ City, University of London\\ EC1V 0HB London, UK}
\emailAdd{Adam.Varga@city.ac.uk}

\abstract{
We obtain explicit formulas for the eight bosonic and eight fermionic 
fluctuations around the mixed-flux generalization of the 
Hofman-Maldacena giant magnon on AdS$_3 \times$S$^3 \times$T$^4$ 
and AdS$_3 \times$S$^3 \times$S$^3 \times$S$^1$.
As a check of our results, we confirm that the semiclassical quantization 
of these fluctuations leads to a vanishing one-loop correction to the 
magnon energy, as expected from symmetry based arguments.
}

\begin{document} 
\maketitle
\flushbottom


\section{Introduction}

An important aspect of the $\AdS_5 / \CFT_4$ correspondence
\cite{Maldacena:1997re} is integrability, a hidden symmetry present 
both on the $\N=4$ super Yang-Mills gauge theory side 
\cite{Minahan:2002ve, Beisert:2003tq, Beisert:2003yb} and 
$\AdS_5 \times \Sphere^5$ type IIB superstring theory side
\cite{Bena:2003wd, Kazakov:2004qf, Arutyunov:2004vx, Beisert:2005bm,
Arutyunov:2004yx} of the duality. Interactions in a quantum integrable
theory reduce to a series of diffractionless two-body scattering processes,
and in the decompactified worldsheet limit the spectrum is solvable 
using a Bethe Ansatz \cite{Bethe:1931hc, Faddeev:1996iy}. Therefore, 
the main object of interest in (the planar limit of) $\AdS / \CFT$ is the 
S-matrix, encoding these two-body scatterings of elementary excitations, 
or magnons \cite{Staudacher:2004tk}. $\AdS_5 / \CFT_4$ has a 
$\psu(2,2|4)$ symmetry, and the subalgebra leaving the vacuum invariant 
is $\su(2|2)^2$. The off-shell, centrally extended version of this residual
algebra, $\ce{\su(2|2)}^2$ fixes the S-matrix up to an overall phase
\cite{Beisert:2005tm, Beisert:2006qh}, which then can be calculated 
from the so-called crossing symmetry \cite{Janik:2006dc, Beisert:2006ib,
Beisert:2006ez, Dorey:2007xn, Volin:2009uv}. These algebraic arguments
also determine the magnon dispersion relation to be 
\begin{equation}\label{fullSpec:IntroMagnonDisp}
\epsilon = \sqrt{ 1 + 4 \hh^2 \sin^2 \frac{\pp}{2} } \ ,
\end{equation}
where $\pp$ is the magnon momentum, and the effective 
string tension $\hh$ is related to the string tension $\alpha'$ and 
$\AdS$ radius $R$ on the string theory side, and the 't Hooft coupling 
$\lambda$ on the gauge theory side by
\begin{equation}
\hh = \frac{R^2}{2\pi\alpha'} =  \frac{\sqrt{\lambda}}{2\pi} \ .
\end{equation}
 
Solitons are particle-like solutions of integrable field theories, 
whose dynamics can be captured by a small number of collective 
degrees of freedom. Quantization of these collective coordinates
\cite{Gervais:1974dc,Gervais:1975pa,Gervais:1975yg,Gervais:1976wr}
provides a window into regimes of the quantum theory not directly 
accessible to perturbation methods. The giant magnon of Hofman and
Maldacena~\cite{Hofman:2006xt} is a soliton of the integrable 
$\AdS_5 \times \Sphere^5$ worldsheet sigma-model~\cite{Metsaev:1998it},
whose spacetime image is an open string uniformly rotating around an 
axis of an $\Sphere^2 \subset \Sphere^5$, stretched between two 
points on the equator. In fact, the worldsheet momentum $\pp$ of the 
giant magnon is the angular distance between these two points, and its
dispersion relation $\epsilon = 2\hh\sin\frac{\pp}{2}$ agrees with the large 
$\hh$ limit of \eqref{fullSpec:IntroMagnonDisp}.
An $\R \times \Sphere^3$ generalization of this solution, the dyonic giant
magnon \cite{Chen:2006gea}, has the dispersion relation
\begin{equation}
\epsilon = \sqrt{ J_2 + 4 \hh^2 \sin^2 \frac{\pp}{2} } \ ,
\end{equation}
where $J_2$ is the second angular momentum on $\Sphere^3$. Upon
semiclassical quantization $J_2$ takes integer values, and we recover the 
exact disperison relation \eqref{fullSpec:IntroMagnonDisp}.

There are a number of calculations one can perform to check that the
giant magnon is indeed the large coupling limit of the elementary
excitation of the quantum theory. A semiclassical analysis of the 
worldsheet scattering of dyonic giant magnons \cite{Chen:2007vs} 
shows that their 1-loop S-matrix agrees with the Hernandez-Lopez phase
\cite{Beisert:2006ib}, and also that the 1-loop correction to the giant 
magnon energy vanishes. From an algebraic perspective the magnon is a 
BPS state of the $\ce{\su(2 | 2)}^2$ superalgebra, and 
accordingly, must be part of a 16 dimensional short multiplet 
\cite{Beisert:2006qh}. As a consequence the giant magnon should have
eight fermionic zero modes, as Hofman and Maldacena argued 
in~\cite{Hofman:2006xt}. These zero modes were explicitly constructed 
by Minahan~\cite{Minahan:2007gf}, starting from the quadratic fermionic 
part of the Green-Schwarz action expanded around the giant magnon.
Quantizing these modes he was also able to reproduce the odd generators 
of the residual algebra. Subsequently, building on Minahan's work, an 
explicit basis of the magnon's fluctuation spectrum was found by
Papathanasiou and Spradlin \cite{Papathanasiou:2007gd}, once again
confirming that the dispersion relation receives no corrections, by showing
that the 1-loop functional determinant vanishes.

With 32 supercharges $\AdS_5/\CFT_4$ has the maximal amount of
supersymmetry possible for any 10 dimensional string theory, but 
integrability has proven to be a powerful tool in understanding other, 
less supersymmetric AdS/CFT dualities too. 
One example is $\AdS_4/\CFT_3$, the duality between ABJM Super 
Chern-Simons and type IIA string theory on $\AdS_4\times CP^3$ with 
24 supersymemtries \cite{Klose:2010ki}, however, for the rest of this 
paper we focus on $\AdS_3/\CFT_2$, and in particular two\footnote{
	There is a third maximally supersymmetric $\AdS_3$
	background, $\AdS_3 \times \Sphere^3 \times {\rm K}3$. 
	It should be possible to apply integrable methods to this background, 
	at least in the orbifold limit of K3, and then it would be interesting 
	to see what the effect of turning on the blow-up modes is.
} backgrounds with maximal supersymmetry allowed for such 
geometries (16 supercharges). One of them is \AdsST, where the radii of 
$\AdS_3$ and $\Sphere^3$ are equal, and the other one is \AdsSSS,
where the $\AdS$ radius $R$ and the radii of the two 3-spheres 
$R_\pm$ satisfy 
\cite{Gauntlett:1998kc}
\begin{equation}
\frac{1}{R_+^2} + \frac{1}{R_-^2} = \frac{1}{R^2}.
\end{equation}
Setting the $\AdS_3$ radius to one, this geometry can be parametrized 
by an angle $\vp$
\begin{equation}
R_{+}^{2} = \frac{1}{\c2}, 
\quad
R_{-}^{2} = \frac{1}{\s2},
\end{equation}
and in fact the $\vp \to 0$ limit covers the \AdsST geometry too, 
once the blown up sphere is compactified on a torus. The type IIB
supergravity equations allow these backgrounds to be supported by 
pure Ramond-Ramond (R-R) flux, pure Neveu–Schwarz-Neveu–Schwarz 
(NS-NS) flux, or mixed R-R and NS-NS fluxes
\begin{align} \label{fullSpec:mixFluxes}
	\begin{split}
		F &\ =\ 2\td{q} \big( \operatorname{Vol}(\AdS_3) 
									+ \cos\vp \operatorname{Vol}(\Sphere^3_+) 
									+ \sin\vp \operatorname{Vol}(\Sphere^3_-) \big) \ ,
		\\[1em] 
		H &\ =\ 2      q  \big( \operatorname{Vol}(\AdS_3) 
									+ \cos\vp \operatorname{Vol}(\Sphere^3_+) 
									+ \sin\vp \operatorname{Vol}(\Sphere^3_-) \big) \ ,
	\end{split}
\end{align}
where $q \in [0,1]$ and $\td{q}=\sqrt{1-q^2}$. 
While the pure NS-NS theory can be solved using a chiral decomposition
\cite{Maldacena:2000hw, Maldacena:2000kv, Maldacena:2001km},
no such method exists when the R-R flux is turned on, instead, it is believed
that the solution will be given in terms of integrable methods\footnote{
	Although it is worth noting that there have been attempts to understand the 
	mixed-flux theory using the hybrid formalism of Berkovits, Vafa and Witten
	\cite{Berkovits:1999im, Ashok:2009jw, Eberhardt:2018exh, Eberhardt:2018vho}.
}, as both
the pure R-R~\cite{Babichenko:2009dk, OhlssonSax:2011ms, Sundin:2012gc} 
and mixed-flux~\cite{Cagnazzo:2012se} theories were shown to be
classically integrable. 

The \AdsST backgrounds with pure R-R and pure NS-NS fluxes arise as 
near horizon limits of the D1/D5 and F1/NS5 brane systems, respectively.
Historically, the mixed-flux background has been thought of as the 
near-horizon limit of bound states of D1/D5- and F1/NS5-branes, but 
it was recently shown that the same worldsheet action arises in a pure 
NS-NS theory with an R-R modulus turned on \cite{OhlssonSax:2018hgc}.
Integrable structures have been identified in the $\CFT_2$ dual to 
\AdsST strings \cite{Sax:2014mea}, and there has also been promising
progress in understanding the $\CFT_2$ dual of string 
theory on \AdsSSS \cite{Boonstra:1998yu, Gukov:2004ym,Tong:2014yna, 
Eberhardt:2017pty, Eberhardt:2018ouy, Eberhardt:2019niq}.

Similarly to the $\AdS_5/\CFT_4$ duality, the symmetry algebra can be 
used to determine both the S-matrix and the all-loop magnon disperison
relation \cite{Borsato:2013qpa, Hoare:2013ida, Lloyd:2014bsa}
\begin{equation}\label{fullSpec:IntroQMagnon}
\epsilon_{\pm} = \sqrt{\left(m \pm q \sqrt{\lambda} \frac{ \pp}{2\pi}\right)^2 
					+ 4\, \qt^2\, \hh^2 \sin^2 \frac{\pp}{2}} \ ,
\end{equation}
where $\hh =  \frac{\sqrt{\lambda}}{2\pi}$ only in the classical string limit, 
and in general $\hh$ will receive quantum corrections. The excitations are 
of mass $m = 1, 0$ for \AdsST and $m=0, \s2, \c2, 1$ for \AdsSSS.
The mixed-flux \AdsST dyonic giant magnon was found by Hoare, 
Stepanchuk and Tseytlin \cite{Hoare:2013lja}, with the dispersion relation
\begin{equation} \label{fullSpec:IntroQGM}
E- J_1 = \sqrt{(J_2 \pm q \hh \pp)^2
							+ 4\, \qt^2\, \hh^2 \sin^2 \frac{\pp}{2}} \ ,
\end{equation}
where $E$ is the spacetime energy and $J_1, J_2$ are two angular momenta
on the $\Sphere^3$. They also noted that upon semiclassical quantization 
$J_2$ takes integer values, and the lowest $J_2=1$ gives an exact match 
to the quantum dispersion relation \eqref{fullSpec:IntroQMagnon}.
Just like in the $\AdS_5$ case, there are a number of semiclassical checks 
on these string solutions. The 1-loop worldsheet S-matrix has been
determined from multi-soliton scattering states in \cite{Stepanchuk:2014kza}
in agreement with the finite-gap calculations of \cite{Babichenko:2014yaa}.
The 1-loop correction to the magnon energy can also be calculated 
from the algebraic curve \cite{Abbott:2012dd}, or directly from the GS action 
\cite{Sundin:2012gc, Sundin:2014ema}. The off-shell residual symmetry
algebras of \AdsST and \AdsSSS are the centrally extended $\psu(1|1)^4$
\cite{Borsato:2013qpa, Borsato:2014hja, Borsato:2014exa, Lloyd:2014bsa} 
and the centrally extended $\su(1|1)^2$ \cite{Borsato:2012ud, 
Borsato:2015mma}, and the BPS magnon must transform in 4 and 2
dimensional short multiplets of these superalgebras, respectively. 
Therefore, the mixed-flux magnon on \AdsST and \AdsSSS should have 4 
and 2 fermion zero modes. We found these zero modes in 
\cite{Varga:2019hqh}, and showed how they can be used to construct 
the odd generators of the residual algebras. Our objective here is 
to find the complete spectrum of fluctuations around the $\AdS_3$ giant
magnon. Throughout, we will only consider the \textit{stationary magnon}, 
a subclass of solutions we identified as the mixed-flux generalisation 
of the HM giant magnon in \cite{Varga:2019hqh}.
The rest of this paper is structured as follows. 

In section~\ref{fullSpec:SecBos} we first review the mixed-flux stationary
giant magnon on \AdsSSS, then write down the spectrum of small bosonic
fluctuations around the classical solution. Although the perturbation
equations are rather complicated, one can construct explicit solutions
algebraically using the dressing method, which we adapt to be more 
suited to the fluctuation analysis. 
In section~\ref{fullSpec:SecFerm} we find the fermionic fluctuations, closely
following the methods developed in \cite{Minahan:2007gf, Varga:2019hqh}
extended to non-zero angular frequencies. Using the symmetries of the
system and an explicit kappa-fixed ansatz, the full $2 \times 32$ component
spinor equations are reduced to a 4 dimensional system, which we can 
solve explicitly.

Finally in section~\ref{fullSpec:SecQuantization} we read off the stability
angles of the fluctuations, and use them to evaluate the 1-loop functional
determinant around the soliton background, following the method of 
Dashen, Hasslacher and Neveu~\cite{Dashen:1975hd}. 
We find that, in agreement with our expectations based on the 
superalgebra, the leading order quantum correction vanishes. 
We conclude in section~\ref{fullSpec:SecConclusion} and present 
some of the lengthier or more technical details in the appendices.

\section{Bosonic sector}
\label{fullSpec:SecBos}

In this section we review the mixed-flux $\AdS_3$ stationary magnon,
and solve for its bosonic fluctuations using a similar approach 
employed to study the $\AdS_5$ magnon in \cite{Papathanasiou:2007gd}. 
We consider the case of the \AdsSSS background in our calculations, 
and comment on how the \AdsST modes can be obtained at the end 
of the section.

The conformal gauge bosonic action can be written in the form
\begin{equation}\label{fullSpec:bosAction}
S = \td{S} [Y] + \frac{1}{\c2}\ S_{+}[X^{+}] + \frac{1}{\s2}\ S_{-}[X^{-}],
\end{equation}
with $\AdS_3$ and $\Sphere_\pm^3$ components
\begin{align}
\begin{split}
\td{S} [Y] &= -\frac{\hh}{2} \int_{\mathcal{M}} \d^2 x \Big[
						\eta^{ab}\p_{a} Y^{i} \p_{b} Y_{i} + \td{\Lambda}\, (Y^2 + 1 ) 
					\Big]  
					\\[0.5em]&\qquad\qquad
					-\frac{\hh q}{3} \int_{\mathcal{B}} \d^3 x\,
						\eps^{abc} \eps_{\mu\nu\rho\sigma}
						Y^{\mu} \p_a Y^{\nu} \p_b Y^{\rho} \p_c Y^{\sigma}
\\[1em]
S_{\pm} [X] &= -\frac{\hh}{2} \int_{\mathcal{M}} \d^2 x\Big[
						\eta^{ab}\p_{a} X^{i}  \p_{b} X_{i} + \Lambda_{\pm}\, (X^2 - 1 ) 
					\Big]  
					\\[0.5em]&\qquad\qquad
					-\frac{\hh q}{3} \int_{\mathcal{B}} \d^3 x \,
						\eps^{abc} \eps_{ijkl}X^{i} \p_a X^{j} \p_b X^{k} \p_c X^{l}
\end{split}
\end{align}
where $\eta^{ab} = \diag(-1,+1)$, the embedding coordinates $Y \in \R^{2,2}$,
$X^{\pm} \in \R^{4}$ are enforced to lie on the unit-radius surfaces
\begin{equation}\label{fullSpec:UnitRadius}
Y^2 = -1, \quad (X^{\pm})^2 = 1
\end{equation}
by the Lagrange multipliers $\td{\Lambda}, \Lambda_{\pm}$, and the 
Wess-Zumino term is defined on a 3d manifold $\mathcal{B}$ such that 
its boundary is the worldsheet $\p \mathcal{B} = \mathcal{M}$. 
The equations of motion
\vspace{1em}
\begin{align}\label{fullSpec:BosEqn}
\begin{split}
(\p^2 - \fixedspaceL{\Lambda_{\pm}}{\td{\Lambda}})\, 
\fixedspaceL{X^{\pm}_{i}}{Y_{\mu}} 
- q\, \fixedspaceL{K^{\pm}_{i}}{\td{K}_{\mu}} &= 0, 
\qquad
\td{K}_{\mu} = \eps^{ab} \eps_{\mu\nu\rho\sigma} 
						Y^{\nu}  \p_a Y^{\rho} \p_b Y^{\sigma},
\\[1em]
(\p^2 - \Lambda_{\pm})\, X^{\pm}_{i} - q\, K^{\pm}_{i} &= 0,
\qquad
 K^{\pm}_{i}  = \eps^{ab}  \eps_{ijkl}X^{\pm}_{j} \p_b X^{\pm}_{k} \p_c X^{\pm}_{l},
\end{split}
\end{align}
need to be supplemented by the conformal gauge Virasoro constraints
\begin{align}
\begin{split}
(\p_0 Y)^2 + (\p_1 Y)^2 
+ \frac{1}{\c2}\left( (\p_0 X^{+})^2 + (\p_1 X^{+})^2 \right) 
\qquad&\\
+ \frac{1}{\s2}\left(  (\p_0 X^{-})^2 + (\p_1 X^{-})^2 \right) &= 0,
\\[1em]
\p_0 Y \cdot \p_1 Y + \frac{1}{\c2} \p_0 X^{+} \cdot \p_1 X^{+}
+ \frac{1}{\s2} \p_0 X^{-} \cdot \p_1 X^{-} &= 0.
\end{split}
\end{align}
Taking scalar products of \eqref{fullSpec:BosEqn} with $Y, X^{\pm}$, 
it follows from \eqref{fullSpec:UnitRadius} and
\begin{equation}
Y^{\mu} \td{K}_{\mu} = 0,
\qquad
X^{\pm\, i} K^{\pm}_i = 0,
\end{equation}
that the Lagrange multipliers take the classical values
\begin{equation}\label{fullSpec:LagrangeMult}
\td{\Lambda} = - Y \cdot \p^2 Y,
\qquad
\Lambda_{\pm} = X^{\pm} \cdot \p^2 X^{\pm}.
\end{equation}

\subsection{The stationary giant magnon}

The classical solution we consider for the rest of this paper is the stationary mixed-flux giant magnon\vspace{-0.5em}
\begin{align}\label{fullSpec:s3s3StationaryMagnon}
\begin{split}
Y^0 + i Y^1  &= e^{it}
\\
X^{-}_{1} + i X^{-}_{2}  & = e^{i \s2\, t}
\\
Z_1 \equiv X^{+}_{1} + i X^{+}_{2}  & =
	e^{i \c2\, t} \left[ \cos\tfrac{\pp}{2} +i \sin\tfrac{\pp}{2}\, \tanh\cY\right]
\\
Z_2 \equiv X^{+}_{3} + i X^{+}_{4}  & =
	e^{-\frac{ i\,  q}{\sqrt{\qt^2-u^2}} \cY} \sin\tfrac{\pp}{2}\, \sech\cY
\end{split}
\end{align}
where the scaled and boosted worldsheet coordinate is 
\begin{equation}\label{fullSpec:magnonYcoord}
\cY = \c2\ \g \sqrt{\qt^2-u^2} \cX,
\qquad
\cX = \g (x - u t)
\end{equation}
and
\begin{equation}
\qt = \sqrt{1-q^2},
\quad
\g^2= \frac{1}{1-u^2}.
\end{equation}
The parameter $u$, restricted to 	$u \in (-\qt,\qt )$, can be regarded as 
the velocity of the magnon. The worldhseet momentum $\pp \in [0,2\pi)$
is not a Noether charge of the action, rather a topological charge of the
soliton, corresponding to the longitudinal distance between the two 
endpoints of the magnon on the equator of $\Sphere_{+}^3$ ($Z_2 =0$). 
The parameters further satisfy
\begin{equation}\label{fullSpec:pRel}
u = \qt \cos\tfrac{\pp}{2}.
\end{equation}

This is  a special case of the dyonic mixed-flux magnon, which was first
constructed in \cite{Hoare:2013lja} for the \AdsST background. 
The stationary magnon was identifed in~\cite{Varga:2019hqh} as the 
mixed-flux equivalent of the Hofman-Maldacena magnon 
\cite{Hofman:2006xt}, as compared to the more general $\AdS_5$ 
dyonic magnon of~\cite{Chen:2006gea}. The dispersion relation\footnote{
	$E$ is the spacetime energy, $J_1$ is the angular momentum
	corresponding to the maximally supersymmetric geodesic 
	along the equators of $\Sphere_{\pm}^{3}$.
}
\begin{equation}
	E - J_1 = 2 \hh \qt\, \sin\tfrac{\pp}{2},
\end{equation}
bears witness to this analogy, to be compared to the similarly simple  
$E - J_1 = 2 \hh \sin\tfrac{\pp}{2}$ for the $q=0$ HM magnon.
The Lagrange multipliers \eqref{fullSpec:LagrangeMult} 
evaluate to the classical values
\begin{equation}
\td{\Lambda} = 1,
\quad
\Lambda_{-} = -\sin^4\!\vp,
\quad
\Lambda_{+} = \cos^4\!\vp \left(1 - 2\, \qt^{-2}\g^2 (\qt^2-u^2)\, \sech^2\cY\right) .
\end{equation}

\subsection{\texorpdfstring{$\AdS_3$}{AdS_3} fluctuation spectrum}

Let us now determine the spectrum of fluctuations around the mixed-flux magnon \eqref{fullSpec:s3s3StationaryMagnon}, starting with the $\AdS_3$ bosons.
We denote the perturbed solution by
\begin{equation}
Y + \delta\, \td{y}
\end{equation}
where $Y$ is the classical solution, $\delta \ll 1$ and the perturbation $\td{y} \in \R^{2,2}$ is bounded. Substituting into the equation \eqref{fullSpec:BosEqn} and expanding to first order in $\delta$ (note that $\td\Lambda$ also receives corrections) we get the perturbation equation
\begin{equation} \label{fullSpec:AdsPertEq}
(\p^2 - 1)\,  \td{y}_{\mu} 
+ (Y\cdot \p^2 \td{y} + q \td{K} \cdot \td{y})\, Y_{\mu}
- q \td{k}_{\mu} = 0
\end{equation}
where $\td{K}_{\mu}$ is as in \eqref{fullSpec:BosEqn} and
\begin{equation}
\td{k}_{\mu} = \eps^{ab} \eps_{\mu\nu\rho\sigma} \left(
						\td{y}^{\nu}  \p_a Y^{\rho} \p_b Y^{\sigma} 
						+ 2 Y^{\nu}  \p_a \td{y}^{\rho} \p_b Y^{\sigma}\right).
\end{equation}
Furthermore, to preserve the norm \eqref{fullSpec:UnitRadius}, the perturbation must be orthogonal to the classical solution
\begin{equation}\label{fullSpec:AdsPertNorm}
Y_{\mu} \td{y}^{\mu} = 0.
\end{equation}

These equations have one massless and two massive solutions.
To get the massless perturbation we make the ansatz
\begin{equation}
\td{y}^{0} = - f \sin t,
\quad
\td{y}^{1} =  f \cos t,
\end{equation}
for which \eqref{fullSpec:AdsPertEq} reduces to the free wave equation
\begin{equation} \label{fullSpec:masslessAdsMode}
\p^2 f =0
\quad\Rightarrow\quad
f = e^{i k x - i \w t}
\end{equation}
satisfying the massless dispersion relation $\w^2 = k^2$. The remaining two massive solutions lie in the transverse directions ($\td{y}^{0} = \td{y}^{1} =  0$) of $\AdS_3$, automatically satisfying \eqref{fullSpec:AdsPertNorm}. A simple plane-wave ansatz gives
\begin{equation} \label{fullSpec:massiveAdsMode}
\td{y}^2 = e^{i k x - i \w t}, 
\quad
\td{y}^3 = \mp i e^{i k x - i \w t},
\qquad
\w^2 = (1 \pm q k)^2  + \qt^2 k^2.
\end{equation}
Note that this is the small $\pp$, fixed $k = \hh \pp$ limit of the mixed-flux
\AdsSSS dispersion relation  \cite{Lloyd:2014bsa}
\begin{equation}
\epsilon_{\pm} = \sqrt{(m \pm  q \hh \pp)^2 
							+ 4\, \qt^2\, \hh^2 \sin^2 \frac{\pp}{2}} \ .
\end{equation}
with mass $m=1$.

\subsection{\texorpdfstring{$\Sphere^3_{-}$}{S^3-} fluctuation spectrum}

The $\Sphere^3_{-}$ fluctuations are very similar to the ones on $\AdS_3$. Substituting the perturbed solution
\begin{equation}
X^{-} + \delta\, \td{x}^{-}
\end{equation}
into \eqref{fullSpec:BosEqn}, we get the first order equations
\begin{equation} \label{fullSpec:SminPertEq}
(\p^2 + \sin^4\!\vp )\,  \td{x}^{-}_{i} 
+ (X^{-}\! \cdot \p^2 \td{x}^{-} - q K^{-}\! \cdot \td{x}^{-})\, X_{i}
- q k^{-}_{i} = 0
\end{equation}
where $K^{-}_{i}$ is as in \eqref{fullSpec:BosEqn} and
\begin{equation}
k^{-}_{i}= \eps^{ab}  \eps_{ijkl} \left(
						 \td{x}^{-}_{j} \p_b X^{-}_{k} \p_c X^{-}_{l}
						+ 2 X^{-}_{j} \p_b\td{x}^{-}_{k} \p_c X^{-}_{l}\right),
\end{equation}
which needs to be supplemented by $X^{-}_{i} \td{x}^{- i} = 0$ 
to preserve the norm. Just like on $\AdS_3$, these equations admit a massless solution
\begin{align}\label{fullSpec:SminSln1}
\begin{split}
 \td{x}^{-}_1 &= - e^{i k x - i \w t} \sin\left(\s2 t\right),
\\
\td{x}^{-}_2 &=  \phantom{-}e^{i k x - i \w t} \cos\left(\s2 t\right),
\qquad \w^2 = k^2,
\end{split}
\end{align}
and two perturbations of mass $m =\s2$\vspace{0.5em}
\begin{align}\label{fullSpec:SminSln2}
\begin{split}
\td{x}^{-}_3 &= \phantom{\mp} e^{i k x - i \w t},
\\
\td{x}^{-}_4 &=  \mp e^{i k x - i \w t} ,
\qquad \w^2 = (\s2 \pm q k)^2  + \qt^2 k^2.
\end{split}
\end{align}

\subsection{\texorpdfstring{$\Sphere^3_{+}$}{S^3+} fluctuation spectrum}

For the $\Sphere^3_{+}$ perturbed solution we write
\begin{equation}
X^{+} + \delta\, \td{x}^{+},
\end{equation}
and also introduce the complex coordinates
\begin{equation}
z_1 =  \td{x}^{+}_1 + i  \td{x}^{+}_2,
\quad
z_2 =  \td{x}^{+}_3 + i  \td{x}^{+}_4,
\end{equation}
so that the perturbed $\Sphere^3_{+}$ component of  \eqref{fullSpec:s3s3StationaryMagnon} can be written as
\begin{equation}
Z_1 + \delta\, z_1,
\quad
Z_2 + \delta\, z_2.
\end{equation}
The equations of motion for the $\Sphere^3_{+}$ fluctuations read
\begin{align} \label{fullSpec:SplusPertEq}
\begin{split}
&\left(\p^2 -\cos^4\!\vp \left(1 - 2\, \qt^{-2}\g^2 (\qt^2-u^2)\, \sech^2\cY\right)  \right)\,  \td{x}^{+}_{i} 
\\[0.5em]&\qquad\qquad\qquad
+ (X^{+}\! \cdot \p^2 \td{x}^{+} - q K^{+}\! \cdot \td{x}^{+})\, X_{i}
- q k^{+}_{i} = 0
\end{split}
\end{align}
where $K^{+}_{i}$ is as in \eqref{fullSpec:BosEqn},
\begin{equation}
k^{+}_{i}= \eps^{ab}  \eps_{ijkl} \left(
						 \td{x}^{+}_{j} \p_b X^{+}_{k} \p_c X^{+}_{l}
						+ 2 X^{+}_{j} \p_b\td{x}^{+}_{k} \p_c X^{+}_{l}\right),
\end{equation}
and to preserve the embedding norm
\begin{equation}\label{fullSpec:SplusNorm}
X^{+}_{i} \td{x}^{+ i} = 0 .
\end{equation}
These equations have two different classes of solutions.

Firstly, there are the zero modes, representing collective coordinates of the magnon. The BMN limit fixes the orientation of the magnon in the $(X^{+}_{1}, X^{+}_{2})$ plane, but there is a rotational freedom in traverse coordinates $(X^{+}_{3}, X^{+}_{4})$ leading to the zero mode
\begin{align}
\begin{split}
z_1 & = 0,
\\
z_2 &= i e^{ - \frac{i\,  q}{\sqrt{\qt^2-u^2}}\cY} \sech\cY.
\end{split}
\end{align}
Furthermore, the magnon breaks the $x$-translation symmetry of the BMN vacuum, leading to the zero mode
\begin{align}
\begin{split}
z_1 & = i e^{i \c2\, t} \sech^2\!\cY,
\\
z_2 &= - e^{ - \frac{i\,  q}{\sqrt{\qt^2-u^2}}\cY} \sech\cY \tanh\cY.
\end{split}
\end{align}
These two normalizable zero modes are presented for completeness, but will not play any further role in our analysis.

The solutions we are interested in are plane-wave fluctuations of the form
\begin{equation}
e^{i k x - i \w t} f(\cY),
\end{equation}
where $ f(\cY)$ is a bounded profile that is stationary in the magnon's frame. 
The equations are too complicated for us to find solutions by substituting 
the plane-wave ansatz into \eqref{fullSpec:SplusPertEq}, we need to look 
for another strategy. The authors of 
\cite{Papathanasiou:2007gd} suggest using the dressing method 
\cite{Zakharov:1973pp, Harnad:1983we, Spradlin:2006wk} to construct 
the scattering state of a magnon and a breather, only then to expand 
this solution in the breather momentum to find the fluctuation as the
subleading term. We find, instead, that it is simpler to apply the 
dressing method to the perturbed BMN vacuum, i.e. the point-like 
string moving along the equator together with fluctuations like
\eqref{fullSpec:SminSln1}--\eqref{fullSpec:SminSln2}, which results 
in the perturbed magnon. The details of this calculation can be found 
in appendix \ref{fullSpec:AppDressing}, here we just present the solutions.
As further confirmation of the validity of  our approach, we show in
appendix~\ref{fullSpec:AppCompareToPS} that applying our 
method in the $\vp=q=0$ limit we recover the expected subset 
of the $\AdS_5 \times \Sphere^5$  fluctuations found in 
\cite{Papathanasiou:2007gd}.

The massles plane-wave solution is given by
\begin{align}\label{fullSpec:masslesSplusMode}
\begin{split}
z_1 & = - i e^{i k x - i \w t} e^{+i \c2\, t}
\Big(\qt k - \w \cos\hp 
\\&\qquad\qquad
- i \sin\hp\, \tanh\cY 
\left(\w - \qt k\, \cosh (\cY + i \hp)\,  \sech\cY \right) \Big) ,
\\[0.5em]
\bar{z}_1 & =\phantom{-} i e^{i k x - i \w t} e^{-i \c2\, t}
\Big(\qt k - \w \cos\tfrac{\pp}{2} 
\\&\qquad\qquad
+ i \sin\tfrac{\pp}{2}\, \tanh\cY 
\left(\w - \qt k\, \cosh (\cY - i \hp)\,  \sech\cY \right) \Big) ,
\\[0.5em]
z_2 &= \phantom{-} i e^{i k x - i \w t}  
\sin\hp\, e^{- \frac{ i\, q}{\sqrt{\qt^2-u^2}}\cY} \sech\cY \left( q k -i \qt k \sin\hp\, \tanh\cY \right),
\\[0.5em]
\bar{z}_2 &= - i e^{i k x - i \w t}  
\sin\hp\, e^{ + \frac{ i\, q}{\sqrt{\qt^2-u^2}}\cY} \sech\cY \left( q k +i \qt k \sin\hp\, \tanh\cY \right),
\end{split}
\end{align}
with 
\begin{equation}
\w^2 = k^2.
\end{equation}
Here $\bar{z}_i$ are not the complex conjugates of $z_i$, rather\footnote{
To preserve the (relative) simplicity of the formulas we consider $\td{x}^{+}_i$ to be complex themselves. Real solutions to \eqref{fullSpec:SplusPertEq} can be readily obtained by taking the real parts of these fluctuations.
}
\begin{equation}
\bar{z}_1 =  \td{x}^{+}_1 - i  \td{x}^{+}_2,
\quad
\bar{z}_2 =  \td{x}^{+}_3 - i  \td{x}^{+}_4,
\end{equation}
The two massive modes both have $m = \c2$. One of them is
\begin{align}\label{fullSpec:bosFluc1}
\begin{split}
z_1 & = - i e^{i k x - i \w t}e^{ \frac{i\,  q}{\sqrt{\qt^2-u^2}}\cY}  e^{+i \c2\, t} 
\sin\hp\, \sech\cY \Big(\w +  \c2+ q k -  \qt k\, \cosh (\cY + i \hp)\, \sech\cY \Big) ,
\\[0.5em]
\bar{z}_1 & =- i e^{i k x - i \w t}e^{ \frac{i\,  q}{\sqrt{\qt^2-u^2}}\cY}  e^{-i \c2\, t} 
\sin\hp\, \sech\cY \Big(\w -  \c2 - q k -  \qt k\, \cosh (\cY - i \hp)\, \sech\cY \Big) ,
\\[0.5em]
z_2 &= \phantom{-}i e^{i k x - i \w t} \Big( \qt k\, \sin^2\!\hp\, \sech^2\!\cY-2 (\qt k - \w \cos\hp ) - 2i( \c2+ q k) \sin\hp\, \tanh\cY \Big)
\\[0.5em]
\bar{z}_2 &=   \phantom{-}i e^{i k x - i \w t} e^{  \frac{2 i\,  q}{\sqrt{\qt^2-u^2}}\cY} \qt k\, \sin^2\!\hp\, \sech^2\!\cY,
\end{split}
\end{align}
with 
\begin{equation}
\w^2 = (\c2 + q k)^2  + \qt^2 k^2,
\end{equation}
while the other one is
\begin{align}\label{fullSpec:bosFluc2}
\begin{split}
z_1 & = - i e^{i k x - i \w t}e^{- \frac{i\,  q}{\sqrt{\qt^2-u^2}}\cY}  e^{+i \c2\, t} 
\sin\hp\, \sech\cY \Big(\w + \c2 - q k -  \qt k\, \cosh (\cY + i \hp)\, \sech\cY \Big) ,
\\[0.5em]
\bar{z}_1 & =- i e^{i k x - i \w t}e^{ - \frac{i\,  q}{\sqrt{\qt^2-u^2}}\cY}  e^{-i \c2\, t} 
\sin\hp\, \sech\cY \Big(\w -  \c2 + q k -  \qt k\, \cosh (\cY - i \hp)\, \sech\cY \Big) ,
\\[0.5em]
z_2 &=   \phantom{-}i e^{i k x - i \w t} e^{-\frac{2 i\, \ q}{\sqrt{\qt^2-u^2}}\cY} \qt k\, \sin^2\!\hp\, \sech^2\!\cY,
\\[0.5em]
\bar{z}_2 &= \phantom{-}i e^{i k x - i \w t} \Big( \qt k\, \sin^2\!\hp\, \sech^2\!\cY-2 (\qt k - \w \cos\hp ) - 2i( \c2- q k) \sin\hp\, \tanh\cY \Big)
\end{split}
\end{align}
with 
\begin{equation}
\w^2 = (\c2 - q k)^2  + \qt^2 k^2.
\end{equation}

\subsection{Bosonic modes in \texorpdfstring{\AdsSSS}{AdS3 x S3 x S3 x S1} string theory}

In addition to the fluctuations we found above, there is of course the massless $\Sphere^1$ mode
\begin{equation}
e^{i k x - i \w t}  
\qquad
\w^2 = k^2.
\end{equation}
However, in a proper quantization of \AdsSSS string theory the sigma-model action \eqref{fullSpec:bosAction} would need to be supplemented by ghosts, cancelling the  massless $\AdS_3$ mode \eqref{fullSpec:masslessAdsMode}, and also a combination of the massless $\Sphere^3_\pm$ modes \eqref{fullSpec:SminSln1}, \eqref{fullSpec:masslesSplusMode}, corresponding to the $\Sphere^3_{+} \times \Sphere^3_{-}$ leg of the BMN geodesic. These are analogous to the longitudinal modes in light-cone gauge, and in our semiclassical analysis we will simply omit them \cite{Frolov:2002av, Park:2005ji}.

In summary, the \AdsSSS magnon has two massless modes (one on the flat $\Sphere^1$ and another one perpendicular to the BMN angle on $\Sphere^3_{+} \times \Sphere^3_{-}$), two $m=1$ fluctuations on $\AdS_3$, two $m=\c2$ modes on $\Sphere^3_{+}$, and two $m=\s2$ modes on $\Sphere^3_{-}$, all with the dispersion relations
\begin{equation}\label{fullSpec:dispRel}
\w^2 = (m \pm q k)^2  + \qt^2 k^2.
\end{equation}

\subsection{Bosonic modes in \texorpdfstring{\AdsST}{AdS3 x S3 x T4} string theory}

Taking the $\vp \to 0$ limit of \AdsSSS blows up the $\Sphere^3_{-}$ factor, which
we can recompactify on a $\Torus^3$ to get the  \AdsST geometry. In this limit the $\AdS_3$ and $\Sphere^1$ fluctuations are unchanged, the $\Sphere^3_{+}$ modes take the same form but become $m=1$, while on $\Sphere^3_{-}$ the massless mode becomes the one unaffected by the ghosts, and the two $m=\s2$ modes become massless $\Torus^4$ modes. In summary, the \AdsST magnon has four massless, and four mass $1$ bosonic fluctuations.

\section{Fermionic sector}
\label{fullSpec:SecFerm}

In this section we solve for the complete fermion fluctuation spectrum
around the mixed-flux stationary magnon \eqref{fullSpec:s3s3StationaryMagnon}.
Our approach will be mostly based on \cite{Varga:2019hqh}, but rather than normalizable 
zero modes, we will be looking for solutions with plane-wave asymptotes.
The leading order (quadratic) action for fermion fluctuations around a general
bosonic string solution $X^\mu(t,x)$ is given by \cite{Cvetic:1999zs}
\begin{equation}\label{fullSpec:FermLagr}
	S_{\text{F}} = \hh \int \d^2x\ \LL_{\text{F}}\ ,
\qquad
	\LL_{\text{F}}= -i\left(\eta^{ab}\delta^{IJ} 
											+  \eps^{ab}\sigma_3^{IJ}\right)\;
 								\bar{\vt}^I\rho_a\cD_b\,\vt^J \ .
\end{equation}
The $\vt^I$ are two ten-dimensional Majorana-Weyl spinors, $\sigma_3^{IJ}$
is the Pauli matrix $\diag(+1,-1)$, and $ \rho_a$ are projections of the ten-dimensional Dirac matrices
\begin{equation}
	\rho_a \equiv e_a^A\, \G_A \ , 
	\qquad 
	e_a^A\equiv \p_a X^\mu E_\mu ^A (X) \ .
\end{equation}
Note the difference in notation compared to the previous section, $X^\mu$ are now
the curved space coordinates of \AdsSSS, and not coordinates of a flat 
embedding space. For the remainder of this section we use Hopf coordinates, 
where the only non-constant components of the stationary magnon are along
$\mu=t,\th^+,\phi_1^+,\phi_2^+,\phi_1^-$ corresponding to the tangent 
space components $A=0,3,4,5,7$, respectively. The covariant derivative is
\begin{equation}
	\cD_a\vt^I =  \big( \delta^{IJ} \big( \p_a+\tfrac{1}{4}\omega_\mu^{AB} \p_a  X^\mu \G_{AB} \big)
									+ \tfrac{1}{48} \sigma_1^{IJ}  \slashed{F}\rho_a 
									+ \tfrac{1}{8} \sigma_3^{IJ} \slashed{H}_a 
							\big) \ \vt^J \ ,
\end{equation}
where $\omega_\mu^{AB}$ is the usual spin-connection,
\begin{align}\label{fullSpec:Hslasha}
	\begin{split}
	\slashed{H}_a  & \equiv\ e_a^A H_{ABC} \G^{BC}
							= \tfrac{1}{6} ( \rho_a \slashed{H}  + \slashed{H} \rho_a)\ ,	
    \end{split}		
\end{align}
and the contracted 3-form fluxes are
\begin{align}\label{fullSpec:slashes}
	\begin{split}
	\slashed{F} &=  12 \td{q} \ \big(\G^{012} + \cos\vp\ \G^{345}
															  	+ \sin\vp\  \G^{678}   \big) \ ,
	\\[1em]
	\slashed{H} &=  12       q  \ \big(\G^{012} + \cos\vp\ \G^{345} 
															  	+ \sin\vp\ \G^{678}    \big)  \ .
	\end{split}
\end{align}

\subsection{The fluctuation equations}

The equations of motion for \eqref{fullSpec:FermLagr} are
\begin{align}
\begin{split}
(\rho_0 + \rho_1)(\cD_0 - \cD_1)\ \vt^1    &=   0 ,
\\[1em]
(\rho_0 - \rho_1)(\cD_0 + \cD_1)\ \vt^2    &=   0 ,
\end{split}
\end{align} 
We proceed by changing variables to the more natural scaled and boosted 
worldsheet coordinates \eqref{fullSpec:magnonYcoord} of the magnon
\begin{equation}\label{fullSpec:boostedCoords}
	\cY = \c2\, \zeta \cX, \quad \cS = \c2\, \zeta \cT, 
	\qquad \zeta =  \g \sqrt{\qt^2-u^2} ,
\end{equation} 
yielding \vspace{-1em}
\begin{align}\label{fullSpec:EomNoZeroMode}
	\begin{split}
		(\rho_0 + \rho_1) \left[ \zeta (1+ u) \g  
												\big( \DD  - \p_\cS \big) \ \vt^1 
												+ \OO \vt^2  \right] &= 0 ,
		\\[1em]
		(\rho_0 - \rho_1) \left[ \zeta (1- u) \g  
												\big( \DDT + \p_\cS \big) \ \vt^2 
												+ \td{\OO} \vt^1 \right] &= 0 .
	\end{split}
\end{align} 
Here we defined the mixing operators
\begin{align} \label{fullSpec:OOdef}
\OO          = -  \frac{1}{48 \c2} \slashed{F} (\rho_0 - \rho_1) \ ,
\qquad
\td{\OO}  =     \frac{1}{48 \c2} \slashed{F} (\rho_0 + \rho_1) \ ,
\end{align}
and fermion derivatives
\begin{align} \label{fullSpec:DDdef}
	\begin{split}
		\DD   &=  \p_\cY + \ha G\ \G_{34} + \ha Q\ \G_{35}  
						-  \frac{(1-u) \g}{48 \c2\, \zeta} 
							\left( \slashed{H} (\rho_0 - \rho_1) 
									+ (\rho_0 - \rho_1) \slashed{H} \right)\ ,
	\\[1em] 
		\DDT &=  \p_\cY + \ha \tilde{G}\ \G_{34}  + \ha Q\ \G_{35} 
						-  \frac{(1+u) \g}{48 \c2\, \zeta} 
							\left( \slashed{H} (\rho_0 + \rho_1) 
									+ (\rho_0 + \rho_1) \slashed{H} \right)\ ,
	\end{split}					
\end{align} 
with\vspace{-1em}
\begin{align}\label{fullSpec:ExplGQ}
	\begin{split}
		G	&= \ \ \frac{\qt^2(1-u) \cosh^2\!\cY - \qt^2 + u^2}
					{\qt \left(\qt^2 \sinh^2\!\cY + u^2\right)} \sech\cY \ ,
	\\
		\tilde{G} &= - \frac{\qt^2(1+u) \cosh^2\!\cY - \qt^2 + u^2}
					{\qt \left(\qt^2 \sinh^2\!\cY + u^2\right)} \sech\cY \ ,
	\\                  
		Q  &= - \frac{q}{\qt \sqrt{\qt^2 - u^2}} \
			 			\sqrt{\qt^2 \sinh^2\!\cY + u^2}\ \sech\cY \ .
	\end{split}
\end{align}

The full Green-Schwarz superstring has a local fermionic symmetry ($\kappa$-symmetry),
that we need to fix for physical solutions. Noting that the operators $(\rho_0 \pm \rho_1)$
are half-rank, nilpotent and commute with the fermion derivatives $\DD$ and $\DDT$,
it is clear that the projectors
\begin{align} \label{fullSpec:KappaProj}
	 K_1 =  \ha \sec\vp\ \Gh^0( \rho_0 + \rho_1) \ ,
 	\qquad  
 	K_2 =  \ha \sec\vp\ \Gh^0( \rho_0 - \rho_1) \ ,
\end{align}
can be used to fix $\kappa$-gauge. Here we introduced a set of ``boosted'' gamma matrices
\begin{equation}\label{fullSpec:newGamma}
	\Gh^0 = \sec\vp \left(  \G^0 - \sin\vp\ \G^7   \right) ,
	\quad
	\Gh^7 = \sec\vp \left(  \G^7 - \sin\vp\ \G^0   \right) ,
	\quad
	\Gh^A = \G^A\  (A\neq 0, 7),
\end{equation}
that simplify the notation in what follows. The kappa-fixed spinors $\Psi^J = K_J \vt^J$
then satisfy\vspace{-0.5em}
\begin{align}\label{fullSpec:EomEquivFixed}
	\begin{split}
		\zeta (1+ u)\g \big( \DD  - \p_\cS \big)  \Psi^1 + K_1 		\OO \Psi^2 	&= 0 ,
	\\[1em] 
		\zeta (1- u)\g \big( \DDT + \p_\cS \big) \Psi^2 + K_2 \td{\OO} \Psi^1	&= 0  .
	\end{split}
\end{align}
Introducing the 6d chirality projector
\begin{equation}
	P_\pm = \ha \left( \Id   \pm  \Gh^{012345} \right) ,
	\qquad
	[ P_\pm , K_J ] = 0  ,
\end{equation}
and, with  $\bar{\rho}_0 = - \Gh_0\, \rho_0\, \Gh_0$, the invertible matrix
\begin{equation}\label{fullSpec:R}
	R = \ha \sec\vp\ \Gh^{012} \left(\bar{\rho}_0 - \rho_0\right),
\end{equation}
we can rewrite the equations, using the boosted gamma matrix basis
\begin{align} \label{fullSpec:EomRDelta}
	\begin{split}
		\zeta (1+ u)\g \big( \DD  - \p_\cS \big) \Psi^1  
			+ \qt \left ( R P_{-}  -  K_1 \Delta\ \Gh^{012} \right) \Psi^2  &= 0 ,
	\\[1em] 
		\zeta (1- u)\g \big( \DDT + \p_\cS \big) \Psi^2
		 	-  \qt \left( R P_{-} -  K_2 \Delta\ \Gh^{012} \right) \Psi^1   &= 0 .
\end{split}
\end{align}
The fermion differential operators are
\begin{align} \label{fullSpec:DDfinal}
\begin{split}
\DD   &=  \p_{\cY} + \ha G\ \Gh_{34} + \ha Q\ \Gh_{35}  
					+ \frac{q (1-u) \g}{\zeta}
						\left(  R P_{-} - (R + \Gh_{12})\,  P_{+}  + \Delta_0\, \Gh_{12} \right) ,
\\[1em] 
\DDT &=  \p_{\cY} + \ha \tilde{G}\ \Gh_{34}  + \ha Q\ \Gh_{35} 
					+ \frac{q (1+u) \g}{\zeta} 
						\left(  R P_{-} - (R + \Gh_{12})\,  P_{+}  + \Delta_0\, \Gh_{12}\right) ,
\end{split}					
\end{align} 
and we define\vspace{-1em}
\begin{align}\label{fullSpec:DeltaDef}
\begin{split}
	&\Delta = - \ha \tan\vp \left( \Gh^{1268} + \Id \right) \G^7 
		  \equiv\ \Delta_0\, \Gh^{0} + \Delta_7\, \Gh^{7} \ ,
\end{split}
\end{align}
with \vspace{-1em}
\begin{align}
	\begin{split}
		\Delta_0 =  - \ha \tan^2\!\vp  \left( \Gh^{1268} + \Id \right)\ ,
		\qquad
		\Delta_7 =  \csc\vp\, \Delta_0\ .
	\end{split}
\end{align}
Note that the only source of structural difference between the equations for 
\AdsSSS and \AdsST is a non-zero $\Delta$, and in fact this was our main
reason to introduce the boosted gamma matrix basis. A much more detailed
derivation of these equations, together with a thorough explanation of 
$\kappa$-gauge fixing, can be found in \cite{Varga:2019hqh}.

\subsection{Ansatz and reduced equations}

To reduce the seemingly complicated \eqref{fullSpec:EomRDelta} to a 
more manageable set of equations we will make an ansatz that reflects the 
symmetries of the system. Firstly, all of $\Gh^{012345}, \Gh^{12}, \Gh^{68}$
commute with the kappa projectors \eqref{fullSpec:KappaProj}, so the 
kappa-fixed spinors can be written as
\begin{equation}\label{fullSpec:KappaFixedAnsatz}
\Psi^{J} =\sum_{\lamP, \lam12, \lambda_{68} \in \{ \pm\}} 
				\mathcal{K}_{J}(\lamP \lam12 ) V_{\lamP, \lam12, \lambda_{68}}^{J}(\cS,\cY),
\end{equation}
where the eigenvalues of $V_{\lamP, \lam12, \lambda_{68}}^{J}$ under $\Gh^{12}, \Gh^{68}$ 
and $\Gh^{012345}$ are $i \lambda_{12}, i \lambda_{68}$ and  $\lambda_P$, respectively.
Note that $\lambda_{12}, \lambda_{68}, \lambda_P$ all take values in $\pm 1$. 
There are multiple ways to make the above ansatz satisfy $K_J \Psi^J = \Psi^J$,
in \cite{Varga:2019hqh} we chose to impose the additional constraint\footnote{
	Note that kappa-fixing reduces the degrees of freedom by half,
	and in our ansatz this is done at the level of the projections
	$\Gh^{34} V^{J} = + i V^{J}$, since $\mathcal{K}_{J}(\lambda)$ are
	invertible.
} $\Gh^{34} V^{J} = + i V^{J}$
and found
\begin{align}
\begin{split}
\mathcal{K}_{1}(\lambda) &=  e^{+i \chi} \sqrt{1 + \lambda Q_{+}\, \sech\cY}  
						- \lambda  e^{-i \chi}\sqrt{1 - \lambda Q_{+}\, \sech\cY}\, \Gh_{45},
\\
\mathcal{K}_{2}(\lambda) &=   e^{+i \td{\chi}} \sqrt{1 - \lambda Q_{-}\, \sech\cY}
						+  \lambda  e^{-i \td{\chi}} \sqrt{1 + \lambda Q_{-}\, \sech\cY}\, \Gh_{45},
\end{split}
\end{align}
where \vspace{-1em}
\begin{equation}
	Q_\pm = \frac{q\sqrt{\qt^2-u^2}}{\qt (1\pm u)} , 
\end{equation}
and \vspace{-1em}
\begin{align} \label{fullSpec:logPhasesq}
	\begin{split}
		\chi (\cY)	    & = \ha \left( 
								\arccot \left( \frac{u \csch \cY}{\qt} \right) 
							- 	\arcsin \left( \frac{\tanh\cY}
													{\sqrt{1-Q_+^2\, \sech^2\!\cY}}
							\right) \right) ,
	\\[1em] 
		\td{\chi} (\cY)	& = \ha \left( 
								\arccot \left( \frac{u \csch \cY}{\qt} \right) 
							+ 	\arcsin \left( \frac{\tanh\cY}
													{\sqrt{1-Q_-^2\, \sech^2\!\cY}}
							\right) \right) .
	\end{split}
\end{align}
While the zero modes are time-independent in the magnon's frame $\p_\cS \Psi^J = 0$, for the $\cS$-dependence of the non-zero modes we make a Fourier ansatz
\begin{equation}
V^{J}(\cS,\cY) = e^{-i \wt \cS} V^{J}(\cY) .
\end{equation}

As opposed to the kappa-projectors, the equations of motion
\eqref{fullSpec:EomRDelta} only commute with $\Gh^{12}$ and $\Gh^{68}$,
and the solutions will not have definite chirality under $\Gh^{012345}$, 
unless $\Delta = 0$, i.e. for the \AdsST background, or on the 
$\Gh^{1268} = -1$ spinor subspace for the \AdsSSS background.
With this in mind, we take the general ansatz
\begin{align} \label{fullSpec:4ansatz}
\begin{split}
\Psi^{J} &=e^{-i \wt \cS}  \left( f_J(\cY)\  \mathcal{K}_{J}(-\lam12)+ 
						 g_J(\cY)\ \mathcal{K}_{J}(\lam12) \Gh_{07} \right) U,
\end{split}
\end{align}
where the constant Weyl\footnote{
	We postpone the analysis of the Majorana condition until later,
	see the discussion around \eqref{fullSpec:Majorana}.
} spinor $U$, that is shared between $\Psi^1$ and $\Psi^2$,
has eigenvalues $i \lambda_{12}, i \lambda_{68}, +i, -1$
under $\Gh^{12}$, $\Gh^{68}$, $\Gh^{34}$, $\Gh^{012345}$, respectively.
The $P_-$ part of the solution is represented by the scalar functions 
$f_1, f_2$, while $g_1, g_2$ correspond to the $P_+$ components.
The validity of such an ansatz is further justified by a quick counting of 
the degrees of freedom. A general Weyls spinor has 16 complex components, 
and after 4 mutually commuting projections, there is a single free component
left, hence we can capture the $\cY$-dependence with a single function $f_J$ multiplying $U$. Substituting \eqref{fullSpec:4ansatz} into
\eqref{fullSpec:EomRDelta}, after a considerable amount of simplification 
we get
\begin{align}\label{fullSpec:RedEom1}
\begin{split}
		& e^{-i \wt \cS} 
		\Bigg[ \bigg( \left(\p_\cY + C_{f_1f_1}\right) f_1 
					+ C_{f_1f_2} f_2 + C_{f_1g_2} g_2 \bigg) \mathcal{K}_{1}(-\lam12)
	\\[0.5em]&\qquad\qquad
				\bigg( \left(\p_\cY + C_{g_1g_1}\right) g_1 
					+ C_{g_1g_2} g_2 + C_{g_1f_2} f_2 \bigg) \mathcal{K}_{1}(\lam12) \Gh_{07}
		\Bigg]\, U = 0 \ ,
\\[1em]
		& e^{-i \wt \cS} 
		\Bigg[ \bigg( \left(\p_\cY + C_{f_2f_2}\right) f_2 
					+ C_{f_2f_1} f_1 + C_{f_2g_1} g_1 \bigg) \mathcal{K}_{2}(-\lam12)
	\\[0.5em]&\qquad\qquad
				\bigg( \left(\p_\cY + C_{g_2g_2}\right) g_2 
					+ C_{g_2g_1} g_1 + C_{g_2f_1} f_1 \bigg) \mathcal{K}_{2}(\lam12) \Gh_{07}
		\Bigg]\, U = 0 \ ,
\end{split}
\end{align}
with coefficients $C_{..}$ listed in appendix \ref{fullSpec:AppRedEqCoeffs}. 
The matrix structure matches that of the general kappa-fixed spinors,  
confirming that the kappa-projectors commute with the fermion 
derivatives $\DD$, $\DDT$. Further substituting
\begin{align}
\begin{split}
f_1 = \frac{1}{\sqrt{1+u}} e^{ i \frac{\lam12 q}{\sqrt{\qt^2 -u^2}}	
											 \left(\tfrac{1}{2} +p_{1268} \tan^2\!\vp \right) \cY}
									e^{ -\frac{i}{2} \lam12
										 \arctan \left( \frac{Q_+\tanh\cY}{\sqrt{1- Q_+^2}}\right)}
									\td{f}_1,
\\
g_1 = \frac{i \lam12} {\sqrt{1+u}} e^{ i \frac{\lam12 q}{\sqrt{\qt^2 -u^2}}	
											 \left(\tfrac{1}{2} +p_{1268} \tan^2\!\vp \right) \cY}
									e^{+ \frac{i}{2} \lam12
										 \arctan \left( \frac{Q_+\tanh\cY}{\sqrt{1- Q_+^2}}\right)}
									\td{g}_1,
\\
f_2 = \frac{\lam12 }{\sqrt{1-u}} e^{ i \frac{\lam12 q}{\sqrt{\qt^2 -u^2}}	
											 \left(\tfrac{1}{2} +p_{1268} \tan^2\!\vp \right) \cY}
									e^{- \frac{i}{2} \lam12 
										 \arctan \left( \frac{Q_-\tanh\cY}{\sqrt{1- Q_-^2}}\right)}
									\td{f}_2,
\\
g_2 = \frac{i}{\sqrt{1-u}} e^{ i \frac{\lam12 q}{\sqrt{\qt^2 -u^2}}	
											 \left(\tfrac{1}{2} +p_{1268} \tan^2\!\vp \right) \cY}
									e^{+ \frac{i}{2} \lam12
										 \arctan \left( \frac{Q_-\tanh\cY}{\sqrt{1- Q_-^2}}\right)}
									\td{g}_2,
\end{split}
\end{align}
where $p_{1268}$ is the eigenvalue of the projector $\tfrac{1}{2}(\Id + \Gh^{1268})$
\begin{equation}
p_{1268} = \tfrac{1}{2}(1-\lam12 \lambda_{68}),
\end{equation}
and defining
\begin{equation}
\xi = \frac{q u}{\sqrt{\qt^2-u^2}} \ ,
\end{equation}
we arrive at the reduced equations
\begin{align}\label{fullSpec:finalReducedEom}
\begin{split}
&\p_\cY \td{f}_1 + i (\wt +(1 + p_{1268} \tan^2\!\vp) \lam12 \xi) \td{f}_1 
\\[0.5em]&\qquad
	+  (1 + p_{1268} \tan^2\!\vp) (\tanh\cY - i \lam12  \xi) \td{f}_2 
	- \lam12\, p_{1268} \tan\vp\sec\vp\, \sech \cY \td{g}_2 = 0,
\\[1em]
&\p_\cY \td{f}_2 - i (\wt + (1 + p_{1268} \tan^2\!\vp)  \lam12 \xi) \td{f}_2
\\[0.5em]&\qquad
	+  (1 + p_{1268} \tan^2\!\vp) (\tanh\cY + i \lam12  \xi) \td{f}_1 
	+ \lam12\, p_{1268} \tan\vp\sec\vp\, \sech \cY \td{g}_1 = 0,
\end{split}
\\[1em]
\begin{split}
&\p_\cY \td{g}_1 + i (\wt + p_{1268} \tan^2\!\vp  \lam12 \xi) \td{g}_1 
\\[0.5em]&\qquad
	+  \lam12\, p_{1268} \tan\vp\sec\vp\, \sech \cY \td{f}_2
	+ p_{1268} \tan^2\!\vp (\tanh\cY + i \lam12  \xi)  \td{g}_2 = 0,
\\[1em]
&\p_\cY \td{g}_2 - i (\wt + p_{1268} \tan^2\!\vp  \lam12 \xi) \td{g}_2
\\[0.5em]&\qquad
	-  \lam12\, p_{1268} \tan\vp\sec\vp\, \sech \cY \td{f}_1
	+ p_{1268} \tan^2\!\vp (\tanh\cY - i \lam12  \xi)  \td{g}_1 = 0.
\end{split}
\end{align}

\subsection{Solutions}

Let us first find the solutions for $\vp >0$, i.e. for the \AdsSSS geometry. 
For $p_{1268}=0$ the $P_{+}$ components $\td{g}_1$ and $\td{g}_2$ decouple, and 
we have the two solutions
\begin{align}
\td{g}_1 &= e^{i \td{k}\cY}, \quad \td{g}_2 = 0, \qquad \td{k} = - \wt,
\\
\td{g}_2 &= e^{i \td{k}\cY}, \quad \td{g}_1 = 0, \qquad \td{k} = + \wt,
\end{align}
while on the $P_{-}$ subspace we have the equations
\begin{align}
\begin{split}
&\p_\cY \td{f}_1 + i (\wt +\lam12 \xi) \td{f}_1 
	+  (\tanh\cY - i \lam12  \xi) \td{f}_2 = 0,
\\[1em]
&\p_\cY \td{f}_2 - i (\wt + \lam12 \xi) \td{f}_2
	+  (\tanh\cY + i \lam12  \xi) \td{f}_1  = 0,
\end{split}
\end{align}
with the two solutions
\begin{align}
\begin{split}
\td{f}_1 &= e^{i \td{k}\cY} \left(\tanh\cY - i(\td{k}-\wt) \right),
\\
\td{f}_2 &= e^{i \td{k}\cY} \left(\tanh\cY - i(\td{k}+\wt) \right),
\\[1em]
\qquad \td{k} &= \pm\sqrt{\wt^2 + 2 \lam12 \xi \wt - 1}.
\end{split}
\end{align}
The case of $p_{1268}=1$ is a bit more complicated, but solving the 
first two equations of \eqref{fullSpec:finalReducedEom} for $\td{g}_J$ and
substituting into the second two, we get two second order differential equations 
for $\td{f}_1, \td{f}_2$. The difference of those two equations is
\begin{equation}
\p_\cY^2 \left( \td{f}_1 - \td{f}_2 \right) 
+ \left( \wt^2 + 2 \lam12 \xi \sec^2\!\vp\, \wt - \sec^4\!\vp  \right) \left( \td{f}_1 - \td{f}_2 \right)  = 0,
\end{equation}
which is easily solved, and inserting the solution into the 
$\left(\td{f}_1 + \td{f}_2\right)$ equation we find
\begin{align}
\begin{split}
\td{f}_1 &= e^{i \td{k}\cY} \lam12 \tan\vp\sec\vp\ \sech\cY,
\qquad
\td{g}_1 = - e^{i \td{k}\cY} \left(\tan^2\!\vp \tanh\cY + i(\td{k}-\wt) \right),
\\
\td{f}_2 &=e^{i \td{k}\cY} \lam12 \tan\vp\sec\vp\ \sech\cY, 
\qquad
\td{g}_2 = \phantom{-}e^{i \td{k}\cY} \left(\tan^2\!\vp \tanh\cY + i(\td{k}+\wt) \right),
\\[0.5em]
\qquad \td{k} &= \pm\sqrt{\wt^2 + 2 \lam12 \xi \tan^2\!\vp \wt - \tan^4\!\vp},
\end{split}
\end{align}
and
\begin{align}
\begin{split}
\td{f}_1 &= e^{i \td{k}\cY} \left(\sec^2\!\vp\, \tanh\cY - i(\td{k}-\wt) \right),
\qquad
\td{g}_1 = \phantom{-}e^{i \td{k}\cY} \lam12 \tan\vp\sec\vp\ \sech\cY,
\\
\td{f}_2 &= e^{i \td{k}\cY} \left(\sec^2\!\vp \tanh\cY - i(\td{k}+\wt) \right),
\qquad
\td{g}_2 = -e^{i \td{k}\cY} \lam12 \tan\vp\sec\vp\ \sech\cY,
\\[0.5em]
\qquad \td{k} &= \pm\sqrt{\wt^2 + 2 \lam12 \xi \sec^2\!\vp\, \wt - \sec^4\!\vp}.
\end{split}
\end{align}

\paragraph{Dispersion relation.}

The observant reader might have already noted that all of the these solutions 
come with a plane-wave factor  $e^{i \td{k} \cY - i \wt \cS}$, satisfying
\begin{equation}\label{fullSpec:almostDispRel}
\td{k}^2 = \wt^2 \pm 2 \xi (\sec^2\!\vp\, m)\, \wt -(\sec^2\!\vp\, m)^2,
\end{equation}
with masses $m=0, \c2, \s2,$ and $1$.
This is not quite the expected dispersion relation, 
and there are two reasons why. Firstly, $(\cS, \cY)$ are scaled versions of the 
boosted worldsheet coordinates $(\cT,\cX)$, but more importantly, the dispersion 
relation \eqref{fullSpec:dispRel} is not relativistically invariant. We therefore 
need to rewrite the fermion fluctuations in the form
\begin{equation}\label{fullSpec:fermOscTransf}
e^{i \td{k} \cY - i \wt \cS} \vartheta(\cY)
=  e^{i (\td{k}+\alpha) \cY - i \wt \cS} e^{-i \alpha \cY}\vartheta(\cY) 
= e^{i k x - i \w t} e^{-i \alpha \cY}\vartheta(\cY) ,
\end{equation}
where $\alpha$ will be necessary to match \eqref{fullSpec:dispRel}.
From \eqref{fullSpec:boostedCoords} it follows that
\begin{equation}
\td{k} = \tfrac{\sec^2\!\vp}{\sqrt{\qt^2-u^2}} (k-u\w) - \alpha,
\qquad
\wt = \tfrac{\sec^2\!\vp}{\sqrt{\qt^2-u^2}} (\w-u k),
\end{equation}
and substituting these into \eqref{fullSpec:almostDispRel} we get the expected relation
\begin{equation}
\w^2 = (m \pm q k)^2  + \qt^2 k^2,
\end{equation}
provided that
\begin{equation}
\alpha = \tfrac{\sec^2\!\vp}{\sqrt{\qt^2-u^2}} \lam12 q\, m.
\end{equation}
Using this transformation we can parametrize the fluctuations by their
wavenumber $k$, and we find that for a given wavenumber there are
two positive frequency, and two negative frequency solutions of each
mass, $m=0, \c2, \s2,$ and $1$.
Further defining
\begin{equation}
\hat{w}_{\pm}=  \frac{1}{2}\arctan \left( \frac{Q_\pm \tanh\cY}
												{\sqrt{1- Q_\pm^2}}\right)  ,
\end{equation}
we collect these solutions below.

\paragraph{Fermion fluctuations with $m=0$.} The massless perturbations are somewhat special, with the positive and negative frequency solutions exciting only one of the two spinors $\Psi^J$. Writing the solutions as
\begin{align}
\begin{split}
\Psi^{J} &=e^{i k x - i \w t}  \hat{g}_J(\cY)\  \mathcal{K}_{J}(\lambda)V_\lambda,
\end{split}
\end{align}
the positive and negative frequency fluctuations are
\begin{align}\label{fullSpec:masslessFinalSln}
\begin{split}
\hat{g}_2  &=  e^{ \frac{1}{2} \frac{i\,  \lambda  q}{\sqrt{\qt^2-u^2}} \cY} e^{ i \lambda \hat{w}_{-}},
\quad \hat{g}_1 = 0, 
\qquad \w = + k,
\\[1em]
\hat{g}_1  &=  e^{ \frac{1}{2} \frac{i\,  \lambda q}{\sqrt{\qt^2-u^2}} \cY} e^{ i \lambda \hat{w}_{+}},
\quad \hat{g}_2 = 0, 
\qquad \w = - k,
\end{split}
\end{align}
and the eigenvalues of the ($k$-dependent) constant Weyl spinor 
$V_\lambda$ under $\Gh^{34}, \Gh^{12}, \Gh^{68}$ and $\Gh^{012345}$
are $+i, i \lambda, i \lambda$ and $+1$, respectively.

\paragraph{Fermion fluctuations with $m=\c2$.} These solutions live on the same subspace as the normalizable zero modes ($\lamP = -1$, $\lam12 \lambda_{68} = 1$) and are given by
\begin{align}
\begin{split}
\Psi^{J} &=e^{i k x - i \w t}  \hat{f}_J(\cY)\  \mathcal{K}_{J}(-\lambda)U_\lambda,
\end{split}
\end{align}
where
\begin{align}\label{fullSpec:cos2FinalSln}
\begin{split}
\hat{f}_1 &=  \frac{1}{\sqrt{1+u}} \left(\tanh\cY 
- i \tfrac{\sec^2\!\vp}{\sqrt{\qt^2-u^2}}\left( (1+u)(k-\w) 
-\lambda q \c2\right)\right)
e^{ -\frac{1}{2} \frac{i\,  \lambda q}{\sqrt{\qt^2-u^2}} \cY} e^{- i \lambda \hat{w}_{+}},
\\
\hat{f}_2 &= \frac{\lambda }{\sqrt{1-u}} \left(\tanh\cY 
- i \tfrac{\sec^2\!\vp}{\sqrt{\qt^2-u^2}}\left( (1-u)(k+\w) 
-\lambda q \c2\right)\right)
e^{ -\frac{1}{2} \frac{i\, \lambda  q}{\sqrt{\qt^2-u^2}} \cY} e^{- i \lambda \hat{w}_{-}},
\\[1em]
w &= \pm\sqrt{(\c2 - \lambda q k)^2  + \qt^2 k^2},
\end{split}
\end{align}
and the ($k$-dependent) constant Weyl spinor $U_\lambda$ has eigenvalues $+i, i \lambda, i \lambda$ and $-1$ under $\Gh^{34}, \Gh^{12}, \Gh^{68}$ and $\Gh^{012345}$, respectively.

\paragraph{Fermion fluctuations with $m=\s2$.}
These fluctuations live on the $\Gh^{1268} = 1$ subspace, and do not have
a definite chirality under $P_{\pm}$ 
\begin{align}
\begin{split}
\Psi^{J} &=e^{i k x - i \w t}  \left( \hat{f}_J(\cY)\  \mathcal{K}_{J}(-\lambda)+ 
						 \hat{g}_J(\cY)\ \mathcal{K}_{J}(\lambda) \Gh_{07} \right) W_\lambda,
\end{split}
\end{align}
\begin{align}\label{fullSpec:sin2FinalSln}
\begin{split}
\hat{f}_1 &= \frac{1} {\sqrt{1+u}} 
\tan\vp \sec\vp\ \sech\cY\
e^{ \frac{1}{2}  \frac{i\, \lambda q}{\sqrt{\qt^2-u^2}} \cY} e^{- i \lambda \hat{w}_{+}},
\\
\hat{f}_2 &=\frac{ \lambda} {\sqrt{1-u}}
\tan\vp \sec\vp\ \sech\cY\
e^{ \frac{1}{2} \frac{i\, \lambda q}{\sqrt{\qt^2-u^2}} \cY} e^{- i \lambda \hat{w}_{-}} 
\\
\hat{g}_1 &= \frac{i} {\sqrt{1+u}} 
\left(\tan^2\!\vp\, \tanh\cY 
+ i \tfrac{\sec^2\!\vp}{\sqrt{\qt^2-u^2}}\left( (1+u)(k-\w) 
-\lambda q \s2\right)\right)
e^{ \frac{1}{2} \frac{i\, \lambda q}{\sqrt{\qt^2-u^2}} \cY} e^{ i \lambda \hat{w}_{+}},
\\
\hat{g}_2 &=\frac{-i \lambda} {\sqrt{1-u}}
\left(\tan^2\!\vp\, \tanh\cY 
+ i \tfrac{\sec^2\!\vp}{\sqrt{\qt^2-u^2}}\left( (1-u)(k+\w) 
-\lambda q \s2\right)\right)
e^{ \frac{1}{2} \frac{i\, \lambda q}{\sqrt{\qt^2-u^2}} \cY} e^{ i \lambda \hat{w}_{-}},
\\[0.5em]
\w &= \pm\sqrt{(\s2 - \lambda q k)^2  + \qt^2 k^2},
\end{split}
\end{align}
and the eigenvalues of the ($k$-dependent) constant Weyl spinor 
$W_\lambda$ under $\Gh^{34}, \Gh^{12}, \Gh^{68}$ and $\Gh^{012345}$
are $+i, i \lambda, -i \lambda$ and $-1$, respectively.

\paragraph{Fermion fluctuations with $m=1$.}
Finally, the heaviest fermions are 
\begin{align}
\begin{split}
\Psi^{J} &=e^{i k x - i \w t}  \left( \hat{f}_J(\cY)\  \mathcal{K}_{J}(-\lambda)+ 
						 \hat{g}_J(\cY)\ \mathcal{K}_{J}(\lambda) \Gh_{07} \right) W_\lambda,
\end{split}
\end{align}
\begin{align}\label{fullSpec:1FinalSln}
\begin{split}
\hat{f}_1 &= \frac{1} {\sqrt{1+u}} 
\left(\sec^2\!\vp\, \tanh\cY 
- i \tfrac{\sec^2\!\vp}{\sqrt{\qt^2-u^2}}\left( (1+u)(k-\w) 
-\lambda q \right)\right)
e^{ -\frac{1}{2} \frac{i\, \lambda q}{\sqrt{\qt^2-u^2}} \cY} e^{- i \lambda \hat{w}_{+}},
\\
\hat{f}_2 &=\frac{ \lambda} {\sqrt{1-u}}
\left(\sec^2\!\vp\, \tanh\cY 
- i \tfrac{\sec^2\!\vp}{\sqrt{\qt^2-u^2}}\left( (1-u)(k+\w) 
-\lambda q \right)\right)
e^{ -\frac{1}{2} \frac{i\, \lambda q}{\sqrt{\qt^2-u^2}} \cY} e^{- i \lambda \hat{w}_{-}} 
\\
\hat{g}_1 &= \frac{-i} {\sqrt{1+u}} 
\tan\vp \sec\vp\ \sech\cY\
e^{- \frac{1}{2}\frac{i\, \lambda q}{\sqrt{\qt^2-u^2}} \cY} e^{ i \lambda \hat{w}_{+}},
\\
\hat{g}_2 &=\frac{i \lambda} {\sqrt{1-u}}
\tan\vp \sec\vp\ \sech\cY\
e^{ -\frac{1}{2}  \frac{i\, \lambda q}{\sqrt{\qt^2-u^2}} \cY} e^{ i \lambda \hat{w}_{-}},
\\[0.5em]
\w &= \pm\sqrt{(1 - \lambda q k)^2  + \qt^2 k^2},
\end{split}
\end{align}
and the constant spinor $W_\lambda$ satisfies the same conditions as for $m=\s2$.

\paragraph{Majorana condition.} In a Majorana basis $(\Gh^A)^* = - \Gh^A$ and the Majorana condition is $(\Psi^J)^* = \Psi^J$.
To impose this condition we need to consider linear combinations of two solutions (from the same mass group) such that the wavenumbers are $k$ and $-k$, the frequencies are of opposite sign (apart from the massless case), and so are the $\lambda$ eigenvalues. Noting that the dispersion relation is invariant under $(k \to -k, \lambda\to-\lambda)$, and
\begin{equation}\label{fullSpec:Majorana}
\mathcal{K}_{1}(\lambda)^* =
 - \lambda \mathcal{K}_{1}(-\lambda) \Gh_{45},
\quad
\mathcal{K}_{2}(\lambda)^* =
  \lambda \mathcal{K}_{2}(-\lambda) \Gh_{45},
\end{equation}
it follows that $(\Psi^J)^* = \Psi^J$ will simply relate 
the constant spinor multipliers of the two components.
We show explicitly how to construct solutions satisfying the Majorana
condition in the massless case. Analogous expressions for the 
massive modes can also be found, but these are quite lengthy. 
Since they do not play any role in the subsequent analysis we do not 
write them explicitly here. We start with the linear combination
\begin{align}
\begin{split}
\Psi^{1} &=
e^{+i k (x+t)}  e^{ +\frac{1}{2} \frac{i\, q}{\sqrt{\qt^2-u^2}} \cY} e^{ +i \hat{w}_{+}}\  \mathcal{K}_{1}(+1)V^1_{+}
\\ &\qquad
+ e^{-i k (x+t)}  e^{- \frac{1}{2} \frac{i\, q}{\sqrt{\qt^2-u^2}} \cY} e^{ -i \hat{w}_{+}}\  \mathcal{K}_{1}(-1)V^2_{-},
\end{split}
\end{align}
where the two components have opposite $k$, $\w$, and $\lambda$.
Its complex conjugate is
\begin{align}
\begin{split}
(\Psi^{1})^* &=
- e^{-i k (x+t)}  e^{ -\frac{1}{2} \frac{i\, q}{\sqrt{\qt^2-u^2}} \cY} e^{ -i \hat{w}_{+}}\  \mathcal{K}_{1}(-1) \Gh_{45} (V^1_{+})^*
\\ &\qquad
+ e^{+i k (x+t)}  e^{\frac{1}{2} \frac{i\, q}{\sqrt{\qt^2-u^2}} \cY} e^{+ i \hat{w}_{+}}\  \mathcal{K}_{1}(+1) \Gh_{45} (V^2_{-})^*,
\end{split}
\end{align}
and $(\Psi^J)^* = \Psi^J$ as long as
\begin{equation}
\Gh_{45} (V^1_{+})^* = - V^2_{-}
\quad\text{and}\quad
\Gh_{45} (V^2_{-})^* =  V^1_{+}.
\end{equation}
These two conditions are equivalent, and consistent with the $\Gh^{34}, \Gh^{12}, \Gh^{68}$ and $\Gh^{012345}$ eigenvalues of $V^1_{+}$ and $V^2_{-}$. We have found an explicit Majorana solution.

\paragraph{Solutions for \AdsST.}

Again, this geometry corresponds to the $\vp \to 0$ limit, the reduced equations \eqref{fullSpec:finalReducedEom} decouple for the $P_{\pm}$ subspaces, and all of the solutions are  the same form as the $p_{1268}=0$ fluctuations above. In particular, we have four massless fermions
\begin{equation}
\begin{gathered}
\Psi^{J} =e^{i k x - i \w t}  \hat{g}_J(\cY)\  \mathcal{K}_{J}(\lambda)V_\lambda,
\\[1em]
\hat{g}_2  =  e^{ \frac{1}{2} \frac{i\,  \lambda  q}{\sqrt{\qt^2-u^2}} \cY} e^{ i \lambda \hat{w}_{-}},
\quad \hat{g}_1 = 0, 
\qquad \w = + k,
\\
\hat{g}_1  =  e^{ \frac{1}{2} \frac{i\,  \lambda q}{\sqrt{\qt^2-u^2}} \cY} e^{ i \lambda \hat{w}_{+}},
\quad \hat{g}_2 = 0, 
\qquad \w = - k,
\end{gathered}
\end{equation}
and four massive fermions\vspace{0.5em}
\begin{equation}\label{fullSpec:massiveT4fermion}
\begin{gathered}
\Psi^{J} =e^{i k x - i \w t}  \hat{f}_J(\cY)\  \mathcal{K}_{J}(-\lambda)U_\lambda,
\qquad
\w = \pm\sqrt{(1 - \lambda q k)^2  + \qt^2 k^2},
\\[1em]
\hat{f}_1 =  \frac{1}{\sqrt{1+u}} \left(\tanh\cY 
- i \tfrac{1}{\sqrt{\qt^2-u^2}}\left( (1+u)(k-\w) 
-\lambda q \right)\right)
e^{ -\frac{1}{2} \frac{i\,  \lambda q}{\sqrt{\qt^2-u^2}} \cY} e^{- i \lambda \hat{w}_{+}},
\\
\hat{f}_2 = \frac{\lambda }{\sqrt{1-u}} \left(\tanh\cY 
- i \tfrac{1}{\sqrt{\qt^2-u^2}}\left( (1-u)(k+\w) 
-\lambda q \right)\right)
e^{ -\frac{1}{2} \frac{i\, \lambda  q}{\sqrt{\qt^2-u^2}} \cY} e^{- i \lambda \hat{w}_{-}},
\end{gathered}
\end{equation}
where under the operators $\Gh^{34}, \Gh^{12}$ and 
$\Gh^{012345}$ the constant spinor $V_\lambda$ has eigenvalues
$+i, i \lambda$ and $+1$, while $U_\lambda$ has eigenvalues
$+i, i \lambda$ and $-1$, respectively.
The difference compared to \eqref{fullSpec:masslessFinalSln},
\eqref{fullSpec:cos2FinalSln} is that the $\Gh^{68}$ eigenvalues of $U_\lambda$, $V_\lambda$ are no longer constrained.

\section{The 1-loop functional determinant}
\label{fullSpec:SecQuantization}

Using the fluctuations found in the previous two sections we now calculate the leading order quantum corrections to the energy of the stationary giant magnon. We follow a similar argument in \cite{Papathanasiou:2007gd}, which is based on well-established quantization techniques for solitons~\cite{Dashen:1975hd,Gervais:1976wr,Gervais:1975yg,Jevicki:1979nr}. By energy we mean the Noether charge combination $E-J_1$, where $E$ is the conserved charge associated with translations in global $\AdS_3$ time, while $J_1$ is the U$(1)$ charge associated with rotations along the BMN geodesic. In light-cone gauge, the quantity $E-J_1$ can be identified with the (transverse) Hamiltonian of physical string excitations~\cite{Frolov:2002av}. 
In conformal gauge the sigma-model action has to be supplemented by ghosts to cancel two unphysical bosons, however, for the purposes of our semiclassical analysis it is sufficient to simply omit two of the massless bosonic modes, as disucssed in Section \ref{fullSpec:SecBos}.

A detailed calculation (building on the \AdsST case discussed in 
\cite{Lloyd:2014bsa}) can be found in \cite{Varga:2019hqh}, here we 
just note that the mixed-flux dyonic giant magnon has classical charges
\begin{equation}
	E - J_1 = \sqrt{\left(\c2\, J_2 - \hh q \pp \right)^2 
						 + 4 \hh^2 \qt^2\, \sin^2\tfrac{\pp}{2}} ,
\end{equation}
where $\c2$ is the mass of the magnon and $J_2$ is its second angular momentum.
Remarkably, this classical expression is in agreement with the exact dispersion relation  of elementary excitations
\begin{equation}
\epsilon = \sqrt{\left(m \pm  q \hh \pp\right)^2 +
						 4\, \qt^2\, \hh^2 \sin^2 \frac{\pp}{2}} \ ,
\end{equation}
determined from 
supersymmetry~\cite{Borsato:2013qpa, Hoare:2013ida, Lloyd:2014bsa},
hence we expect no quantum corrections. The one-loop correction to 
the energy can be calculated as the functional determinant 
$\ln \det|\delta^2 S|$ around the classical background, and is given by 
\begin{equation}
\label{fullSpec:1loopCorrection}
\frac{1}{2} \sum_{i, k} (-1)^F \nu_i,
\end{equation}
where $F$ is the fermion number operator, $\nu_i$ are the so-called 
stability angles, frequencies of small oscillations around the classical solution, and the sum is over excitations $i$ and wavenumbers $k$.
For a non-static soliton, like the giant magnon, we can apply the method of 
Dashen, Hasslacher and Neveu~\cite{Dashen:1975hd} to calculate these stability angles. We put the system in a box of length $L \gg 1$, with periodic boundary conditions $x \cong x + L$. It is clear from the form of the solution
\eqref{fullSpec:s3s3StationaryMagnon} that the system is also periodic 
in worldsheet time, with period $T = L/u$. Then, the stability angle $\nu$ of a generic 
fluctuation $\delta \phi$ can be read off from
\begin{equation}
\delta \phi(t + T, x) = e^{-i \nu} \delta \phi(t,x).
\end{equation}

Although we had to write the oscillations in the original worldsheet coordinates 
$(x,t)$ to get the correct dispersion relations, the magnon's stationary frame 
$(\cX,\cT)$ is better suited to the analysis of stability angles. In Sections
\ref{fullSpec:SecBos} and \ref{fullSpec:SecFerm} we found fluctuations with
oscillatory terms
\begin{equation}
e^{i k x - i \w t}
\end{equation}
parametrized by mass $m$ and an additional eigenvalue $\lambda = \pm 1$, and satisfying
dispersion relations
\begin{equation}
\w= \sqrt{(m - \lambda q k)^2  + \qt^2 k^2}.
\end{equation}
Rewriting the plane-wave terms as\footnote{
	Note that this is the inverse of the transformation 
	\eqref{fullSpec:fermOscTransf} that we applied to the fermion fluctuations.
}
\begin{equation}\label{fullSpec:KWinXT}
e^{i k x - i \w t} = e^{i \hat{k} \cX - i \hat{\w} \cT} 
						 e^{i \lambda q m \g \cX},
\end{equation}
the new frequency and wavenumber satisfy
\begin{equation}
\hat{\w}= -\lambda q u \g m +  \sqrt{ \qt^2 m^2 + \hat{k}^2},
\end{equation}
while $e^{i \lambda q m \g \cX}$ can be absorbed into
the rest of the $\cY$-dependent solution.

\subsection{1-loop correction in \texorpdfstring{\AdsSSS}{AdS3 x S3 x S3 x S1} string theory}

For each excitation, the stability angle can be further decomposed as
\begin{equation}
\nu_i(m,\lambda) = \nu^{(0)}_i(m,\lambda) 
							+ \nu^{(1)}_i(m,\lambda) + \lambda \nu^{(2)}_i(m),
\end{equation}
where $\nu^{(0)}_i$ comes from the pure plane-wave $
e^{i \hat{k} \cX - i \hat{\w} \cT}$, $\nu^{(2)}_i$ from terms like $e^{i\lambda f(\cY)}$ and  $\nu^{(1)}_i$ corresponds to the rest. Since we have exactly one boson and one fermion for each of the 8 combinations of $(m,\lambda)$, and the first terms are the same
\begin{equation}
 \nu^{(0)}_{bos}(m,\lambda) =  \nu^{(0)}_{ferm}(m,\lambda)
 = \frac{L}{u}\, \g \left(\hat{\w} + u \hat{k} \right),
\end{equation}
the total contribution form these terms vanishes even before integrating over $\hat{k}$
\begin{equation}
\sum_{m,\lambda}  \nu^{(0)}_{bos}(m,\lambda) - \nu^{(0)}_{ferm}(m,\lambda) = 0.
\end{equation}
Furthermore, summing over $\lambda =\pm 1$ pairs of the same excitation the $\nu_i^{(2)}$ terms cancel, leaving us with the total correction
\begin{equation} \label{fullSpec:1loopIntegral}
 \sum_{i, k} (-1)^F \nu_i = \int \d \hat{k} \sum_{m,\lambda} \left( 
 			\nu^{(1)}_{bos}(m,\lambda) - \nu^{(1)}_{ferm}(m,\lambda)\right).
\end{equation}
Under the transformation \eqref{fullSpec:KWinXT} we have
\begin{equation}
k = \g (\hat{k} + u \hat{\w}) + \lambda q \g^2 m ,
\qquad
\w = \g (\hat{\w} + u \hat{k}) + \lambda q u \g^2 m ,
\end{equation}
and it is then straightforward to read off the $\nu^{(1)}_i$ stability angles for the fluctuations in Sections \ref{fullSpec:SecBos} and \ref{fullSpec:SecFerm}. The excitations with non-zero $\nu^{(1)}_i$ are the two $m=\c2$ bosons \eqref{fullSpec:bosFluc1},  \eqref{fullSpec:bosFluc2} 
with
\begin{equation}
e^{\nu^{(1)}_{bos}(\c2,\lambda)} = E_{bos}(\c2 ,\lambda),
\end{equation}
and six massive fermions  \eqref{fullSpec:cos2FinalSln}, \eqref{fullSpec:sin2FinalSln}, \eqref{fullSpec:1FinalSln} with
\begin{align}
\begin{split}
e^{\nu^{(1)}_{ferm}(\c2,\lambda)} &= E_{ferm}(\c2 ,\lambda),
\\
e^{\nu^{(1)}_{ferm}(\s2,\lambda)} &= 1/E_{ferm}(\s2 ,\lambda),
\\
e^{\nu^{(1)}_{ferm}(1,\lambda)} &= E_{ferm}(1,\lambda),
\end{split}
\end{align}
where we have defined
\begin{align}
\begin{split}
E_{bos}(m,\lambda) &= \
\frac{\hat{k} - \tfrac{q^2 u}{\qt^2-u^2}\, \hat{\w} + \lambda q \g m +
         i\left( \g\sqrt{\qt^2-u^2} m - \tfrac{\lambda q}{\sqrt{\qt^2-u^2}}\, 
         		(\hat{k}+u\hat{\w})\right)}
         {\hat{k} - \tfrac{q^2 u}{\qt^2-u^2}\, \hat{\w} + \lambda q \g m -
         i\left( \g\sqrt{\qt^2-u^2} m - \tfrac{\lambda q}{\sqrt{\qt^2-u^2}}\, 
         		(\hat{k}+u\hat{\w})\right)},
\\[1em]
E_{ferm}(m,\lambda) &= \
\frac{\hat{k} - \hat{\w} + i \g\sqrt{\qt^2-u^2} m}
        {\hat{k} - \hat{\w} - i \g\sqrt{\qt^2-u^2} m}.
\end{split}
\end{align}
With these, the integrand of \eqref{fullSpec:1loopIntegral} becomes
\begin{equation}\label{fullSpec:stabAnglesS1}
\sum_{m,\lambda} \left(\nu^{(1)}_{bos}(m,\lambda) 
									- \nu^{(1)}_{ferm}(m,\lambda)\right) = 
-i \log\left(\prod_{\lambda =\pm 1}
	 \frac{E_{bos}(\c2,\lambda) E_{ferm}(\s2 ,\lambda)}
	 		{E_{ferm}(\c2 ,\lambda) E_{ferm}(1 ,\lambda)}\right).
\end{equation}
Since
\begin{equation}\label{fullSpec:stabilityAngleIdentity}
	 \frac{E_{bos}(m,+1)E_{bos}(m,-1)}
	 		{\left(E_{ferm}(m,+1)E_{ferm}(m,-1)\right)^2} = 1
\end{equation}
holds for general $m$, \eqref{fullSpec:1loopIntegral} simplifies to
\begin{equation}\label{fullSpec:1loopIntegral2}
\sum_{i, k} (-1)^F \nu_i =-i  \int \d \hat{k} \log\left(\prod_{\lambda =\pm 1}
	 \frac{E_{ferm}(\c2,\lambda) E_{ferm}(\s2 ,\lambda)}
	 		{E_{ferm}(1 ,\lambda)}\right).
\end{equation}
Further noting that
\begin{equation}
	E_{ferm}(m,+1)E_{ferm}(m,-1) = 
		\frac{\hat{k} + i \g\sqrt{\qt^2-u^2} m}{\hat{k}  - i \g\sqrt{\qt^2-u^2} m}
\end{equation}
it is clear that the integrand is antisymmetric in $\hat{k}$.  
Moreover, we have the asymptotic expansion around $\hat{k} = \pm \infty$
\begin{equation}
\prod_{\lambda =\pm 1}
	 \frac{E_{ferm}(\c2,\lambda) E_{ferm}(\s2 ,\lambda)}
	 		{E_{ferm}(1 ,\lambda)} = 1
	 		 + \frac{i}{2} \g^3 (\qt^2-u^2)^{3/2}\sin^2\!2\vp\ \frac{1}{\hat{k}^3} 
	 		 + O\!\left(  \frac{1}{\hat{k}^5} \right),
\end{equation}
and taking logarithm, the integrand of \eqref{fullSpec:1loopIntegral2} 
is $O\!\left(\hat{k}^{-3}\right)$, hence the integral itself is bounded 
and well-defined. We conclude that the integral is zero, and, in agreement
with our expectations, the giant magnon energy receives no corrections 
at one loop, providing another check on our results.

\subsection{1-loop correction in \texorpdfstring{\AdsST}{AdS3 x S3 x T4} string theory}

On \AdsST the situation is even simpler. We have two bosons and two
fermions for each of the 4 combinations of $m=0,1$, $\lambda=\pm1$. 
Paring these up, the $\nu^{(0)}_i$ contribution vanishes, while the 
$\nu^{(2)}_i$ terms cancel between $\lambda=\pm1$ pairs. With two 
of the massive bosons and four of the massive fermions contributing, 
the integrand of \eqref{fullSpec:1loopIntegral} becomes
\begin{equation}
\sum_{\lambda} \left(\nu^{(1)}_{bos}(1,\lambda) 
									- 2 \nu^{(1)}_{ferm}(1,\lambda)\right) = 
-i \log\left(
	 \frac{E_{bos}(1,+1)E_{bos}(1,-1)}
	 		{\left(E_{ferm}(1,+1)E_{ferm}(1,-1)\right)^2} \right),
\end{equation}
which is the same as the $\vp \to 0$ limit of \eqref{fullSpec:stabAnglesS1}.
Using \eqref{fullSpec:stabilityAngleIdentity} we arrive at the expected
zero one-loop correction result even before integrating over $\hat{k}$.

\section{Conclusions}
\label{fullSpec:SecConclusion}

In this paper we found the full spectrum of fluctuations around the 
mixed-flux $\AdS_3$ stationary giant magnon, the $q>0$ generalisation
of the Hofman-Maldacena giant magnon. To obtain the non-trivial 
bosonic fluctuations, we adapted the method used in 
\cite{Papathanasiou:2007gd}. Rather than dressing the vacuum 
twice to get a complicated breather-soliton superposition (only then 
to expand in small breather momentum), we dress the perturbed BMN
vacuum once, keeping terms up to subleading order throughout the
calculation. The leading order term in the dressed solution is the
giant magnon, so the subleading term must be its perturbation.
The fermionic fluctuations are obtained as solutions of the equations 
derived from the quadratic fermionic action, using the formalism
developed in \cite{Varga:2019hqh}, which builds on the original 
developments of \cite{Minahan:2007gf} for $\AdS_5$.

We find that all of the fluctuations can be written in the form
\begin{equation}
e^{i k x - i \w t} f(x-ut)
\end{equation}
where $u$ is the magnon's speed on the worldsheet \eqref{fullSpec:pRel}, 
and with $k, \w$ satisfying 
\begin{equation}
\w^2= (m \pm q k)^2  + \qt^2 k^2,
\end{equation}
which is the small-momentum limit of the exact dispersion 
relation~\eqref{fullSpec:IntroQMagnon}. Furthermore, the 
fluctuations can be arranged into short multiplets of the residual 
symmetry algebras, according to mass and chirality ($\pm$ sign in the
dispersion relation). On \AdsST there are four 4 dimensional multiplets of 
$\ce{\psu(1|1)^4}$ with two bosons and two fermions, while \AdsSSS 
has eight 2 dimensional multiplets of $\ce{\su(1|1)^2}$, with a boson 
and a fermion each.

Finally, from the explicit form of each fluctuation we read off the so called
stability angles, which sum to the one-loop functional determinant. 
In both of the geometries we were able to show that this one-loop
determinant is zero, or in other words, the one-loop correction to the 
magnon energy vanishes. It is interesting to compare this result with other
calculations of the one-loop correction to energies of $\AdS_3$ string 
states. The expansion of the coupling $\hh$ around the classical string limit
\begin{equation}
\hh(\lambda) =  \frac{\sqrt{\lambda}}{2\pi} + c 
						+ \mathcal{O} \left( \frac{1}{\sqrt{\lambda}}\right),
\end{equation}
is equivalent to the expansion of the energy \eqref{fullSpec:IntroQMagnon}
\begin{equation}
\epsilon = \epsilon_{0} + \frac{4\, \qt^2\, \hh_{0}^2 \sin^2\!\hp}
											 { \hh_{0} \epsilon_{0} } c  
				+ \mathcal{O} \left( \frac{1}{\sqrt{\lambda}}\right),
\end{equation}
where the subscript $0$ refers to the classical (string) values, and we 
see that our results translate to $c = 0$ for both geometries. 
The one-loop correction to the giant magnon energy on \AdsSSS with 
pure R-R flux was derived in \cite{Sundin:2012gc} directly from the 
GS action, and in \cite{Abbott:2012dd} from the algebraic curve. 
They both found that the correction is dependent on the chosen
regularisation scheme, with two naturally emerging prescriptions: 
in the \textit{physical} regularisation the cutoff is at the same mode 
number for all excitations, while in the \textit{new} prescription the 
cutoff is proportional to the mass of the polarisation. The two 
prescriptions both give zero correction $c_{phys} = c_{new} = 0$ 
on the \AdsST background, but differ for the \AdsSSS theory
\begin{equation}
c_{phys} = \frac{\alpha\log\alpha + (1-\alpha)\log(1-\alpha)}{2\pi}, 
\qquad
c_{new} = 0.
\end{equation}
For the mixed-flux \AdsST background the direct GS action calculation \cite{Sundin:2014ema} shows that there is no one-loop correction, 
$c=0$, and the same conclusion can be drawn by considering the 
worldsheet scattering of giant magnons \cite{Stepanchuk:2014kza}. 
Our results are in agreement with the \textit{new} prescription, although 
it is not clear that we work in either of the regularisation schemes, 
as in \eqref{fullSpec:1loopIntegral2} we have an implicit cutoff\footnote{
	The integrals should be computed separately for each mass before
	summing, instead we first sum, then compute the integral, which is
	 equivalent to having the same cutoff on $\hat{k}$ for each mass.
} on the mode numbers $\hat{k}$ in the magnon's frame.
There have been recent advances in our understanding of the 
protected spectrum of $\AdS_3/\CFT_2$ using integrable methods
\cite{Borsato:2016kbm, Baggio:2017kza}. The protected spectrum for 
\AdsST agrees with the older results of \cite{deBoer:1998kjm}, while 
the \AdsSSS case was independently derived using supergravity and 
WZW methods in \cite{Eberhardt:2017fsi}.

As we have discussed in \cite{Varga:2019hqh}, there is no 
stationary magnon for $q=1$, hence the results of present paper 
do not apply in this limit. The pure NS-NS string theory has been 
long known to be solvable using a chiral decomposition 
\cite{Maldacena:2000hw, Maldacena:2000kv, Maldacena:2001km}, 
and there is now a good understanding of integrability for the microscopic 
excitations \cite{Baggio:2018gct, Dei:2018mfl,Dei:2018yth, Dei:2018jyj}, 
but it would be interesting to see a soliton analysis on these backgrounds.
In more recent developments, the $\CFT$ dual of the $k=1$ WZW model, 
i.e. \AdsST with minimal quantized NS-NS flux, has been identified as a 
symmetric product orbifold 
\cite{Giribet:2018ada, Gaberdiel:2018rqv,Eberhardt:2018ouy, Eberhardt:2020akk}. 

Semiclassical methods continue to provide valuable insight into the
$\AdS_3 / \CFT_2$ duality, as one can see in this paper, the analysis
of fermion zero modes \cite{Varga:2019hqh}, or the calculation of 1-loop
corrections to the rigid spinning string dispersion relations 
\cite{Nieto:2018jzi}. Where they seem to fail is the description of 
massless modes. In the $\alpha \to 0$ limit our fluctuations simply reduce to 
the plane-wave perturbations of the BMN vacuum, shedding no further 
light on the nature of the massless soliton of the theory, in agreement with
our previous findings \cite{Varga:2019hqh}, and the fact that the 
$\alpha \to 0$ limit of the spin-chain fails to capture these 
inherently non-perturbative modes on the other side of the 
duality~\cite{Sax:2012jv}. Furthermore, massless modes render a
perturbative computation of wrapping corrections impossible, once
the theory is put on a compactified worldsheet \cite{Abbott:2015pps}.
Instead, wrapping corrections may be computed from a 
non-perturbative TBA using an alternative low-momentum expansion
\cite{Bombardelli:2018jkj, Fontanella:2019ury, Abbott:2020jaa}.

There are two interesting directions for future research. Firstly, we would 
like to get a better understanding of the solitons of the $q=1$ theory, 
their zero modes and fluctuation spectrum. Secondly, and this is a more
speculative direction, one could try and describe the elusive massless modes
by finding solitons of the fermionic part of the action, motivated by the fact
that the massless representation's highest weight state is a fermion \cite{Borsato:2014hja}.

\acknowledgments

AV would like to thank B. Stefanski for useful discussions. 
AV acknowledges the support of the George Daniels postgraduate scholarship.

\appendix

\section{Dressing the perturbed BMN string}
\label{fullSpec:AppDressing}

In this section we briefly review the dressing method for the $SU(2)$ principal chiral model \cite{Zakharov:1973pp, Harnad:1983we, Spradlin:2006wk} , and describe how it can be used when working with perturbations. We then apply this recipe to the three different perturbed BMN strings to obtain the three $\Sphere^3$ fluctuations of the \AdsST giant magnon. The $\Sphere^3_{+}$ perturbation of the \AdsSSS magnon can be obtained from these, simply by scaling the worldsheet coordinates by $\c2$.

\subsection{Review of the \texorpdfstring{$\SU(2)$}{SU(2)} dressing method}

The $\Sphere^3$ part of the sigma-model action \eqref{fullSpec:bosAction} is equivalent to the $\SU(2)$ principal chiral model with Wess-Zumino term
\begin{equation}\label{fullSpec:PCMAction}
S = -\frac{\hh}{2} \Big[ \int_{\mathcal{M}} \d^2 x\,\tfrac{1}{2}\tr(\bar \Jfrak\, \Jfrak) 
					- q\int_{\mathcal{B}}\d^3x\,\tfrac{1}{3}\varepsilon^{abc}\tr(\Jfrak_a\Jfrak_b\Jfrak_c) \Big]\ .
\end{equation}
where the left currents are $\Jfrak_a = g^{-1}\p_a g$, the partial derivatives on  $\Jfrak = g^{-1} \p g$, $\bar \Jfrak = g^{-1} \pbar g$ are with respect to $z = \ha (t - x)$ and $\zbar = \ha (t+x)$, and the embedding 
\begin{equation}
g  = \begin{pmatrix}
		              Z_1     &  -i Z_2       \\
				    -i \Zbar_2 & \Zbar_1  \\
				\end{pmatrix} \in  \SU(2)
\end{equation}
connects to the $\R^4$ coordinates of \eqref{fullSpec:bosAction}
\begin{align}
\begin{split}
Z_1 &= X_1 + i X_2, \qquad \Zbar_1 = X_1 - i X_2,
\\[1em]
Z_2 &= X_3 + i X_4, \qquad \Zbar_2 = X_3 - i X_4.
\end{split}
\end{align}
Note that $\Zbar_i$ are the complex conjugates of $Z_i$ for the real classical solution, but not necessarily for the perturbation that we will write as complex functions.

The equations of motion for the action \eqref{fullSpec:PCMAction} are
\begin{equation}\label{fullSpec:PCeom}
	(1+q)\pbar \left( \p g \, g^{-1} \right) + 
	(1-q)\p \left( \pbar g \, g^{-1} \right)  = 0,
\end{equation}
and starting with a solution $g$, the dressing method aims to find the appropriate dressing factor factor $\chi(z,\zbar)$ such that 
\begin{equation}\label{fullSpec:gauge}
	g \to g' = \chi g
\end{equation}
is a new solution. The construction exploits the equivalence between the compatibility condition of the overdetermined \textit{auxiliary system}
\begin{equation}\label{fullSpec:aux}
	\pbar \Psi = \frac{A \Psi}{1 + (1+q)\lambda}, \qquad
	\p 	\Psi   = \frac{B \Psi }{1 - (1-q)\lambda},
\end{equation}
and \eqref{fullSpec:PCeom} via
\begin{equation}\label{fullSpec:AB}
	A = \pbar g\,g^{-1}, 
	\qquad 
	B = \p g\,g^{-1}.
\end{equation}
One then solves the auxiliary problem for general complex spectral parameter $\lambda$ such that $\Psi(\lambda)$ satisfies
\begin{equation}\label{fullSpec:dressInitial}
\Psi(0) = g,
\end{equation}
and the unitarity condition\vspace{-0.5em}
\begin{equation}\label{fullSpec:dressPsiUn}
\Psi^\dagger(\bar{\lambda}) \Psi(\lambda) = 1 .
\end{equation}
The simplest non-trivial dressing factor is then given by
\begin{equation}\label{fullSpec:dressChi}
	\chi(\lambda) = \1 + \frac{\lambda_1 - \bar{\lambda}_1}
										  { \lambda     -         \lambda_1}  P,
\end{equation}
with the projector
\begin{equation}\label{fullSpec:dressProjector}
	P = \frac{v_1 v_1^\dagger}{v_1^\dagger v_1} ,
	\quad
	v_1 = \Psi(\bar{\lambda}_1) e,
	\quad
	e = (1, 1).
\end{equation}
We will also refer to the matrix $X$ and scalar $y$
\begin{equation}
X = v_1 v_1^\dagger, \quad y = v_1^\dagger v_1
\quad : \quad
P = \frac{X}{y}.
\end{equation}
In order for the dressed solution $\chi(0) \Psi(0)$  to have unit determinant, we need to introduce an additional constant phase $(\lambda_1/\bar{\lambda}_1)^{1/2}$, and with this, the dressed solution becomes
\begin{equation}\label{fullSpec:dressedSln}
	g' = \sqrt{\frac{\lambda_1}{\bar{\lambda}_1}}
			\left( \Id - \left(1 - \frac{\bar{\lambda}_1}{\lambda_1}\right) P  \right) g\ .
\end{equation}

\subsection{Dressing the unperturbed BMN string}

To set the scene and some notation, let us quickly run through the application of the dressing method to the BMN string $Z_1 = e^{i t},	Z_2 = 0$. We solve the auxiliary problem
\begin{equation}
	g_{\rm{BMN}} = 	\begin{pmatrix}
					e^{-i (z-\bar{z})} 	&  0							\\
					0								&  e^{i (z-\bar{z})}	\\
				\end{pmatrix},
	\qquad
	A_{\rm{BMN}} = - B_{\rm{BMN}} = \begin{pmatrix}
								-1 & 0 \\
								 0 & 1 \\
							\end{pmatrix},
\end{equation}
to find
\begin{equation}\label{fullSpec:PsiVac}
	\Psi_{\rm{BMN}}(\lambda) = 
				\begin{pmatrix}
					e^{-i Z(\lambda)} 	&  0							\\
					0								&  e^{i Z(\lambda)}	\\
				\end{pmatrix},
	\qquad
	Z(\lambda)  = 	\frac{z}{1 - (1 -q)\lambda} 
							- 	\frac{\zbar}{1+ (1+q)\lambda}.
\end{equation}
Introducing the real variables
\begin{equation} \label{fullSpec:UV_Z}
	U =  i \left( Z(\bar{\lambda}_1) - Z(\lambda_1) \right) ,
	\qquad
	V = - Z(\bar{\lambda}_1) - Z(\lambda_1) - t ,
\end{equation}
the projector \eqref{fullSpec:dressProjector} becomes
\begin{equation}
P_{\rm{BMN}} = \frac{X_{\rm{BMN}}}{y_{\rm{BMN}}}
\quad : \qquad
y_{\rm{BMN}} = 2 \cosh U, \quad
X_{\rm{BMN}}  = \begin{pmatrix}
				e^{-U}					&  e^{i  (t + V)}	\\
				 e^{-i  (t + V)}	& 	e^{U}	
			\end{pmatrix}.
\end{equation}
Pametrizing the pole as 
$
\lambda_1 =r e^{i \hp}  ,
$
the dressing \eqref{fullSpec:dressedSln} yields the giant magnon
\begin{equation}
g_{\rm{GM}}  = 
\begin{pmatrix}
	e^{i t} \left[ \cos \tfrac{p}{2} + i \sin \tfrac{p}{2} \, \tanh U \right]    
&  
	-i e^{i V} \sin \tfrac{p}{2} \ \sech{U}     
\\
	-i e^{-i V} \sin \tfrac{p}{2} \ \sech{U}
& 
	e^{-i t} \left[ \cos \tfrac{p}{2} - i \sin \tfrac{p}{2} \, \tanh U \right] 
\\ \end{pmatrix}.
\end{equation}
Furthermore, setting $r = \qt^{-1}$, we get the stationary magnon
\begin{equation}
U = \g \sqrt{\qt^2-u^2}\, \cX ,
\quad
V 	=  - q \g\,  \cX,
\qquad
\cX = \g ( x - u t), 
\end{equation}
where
\begin{equation}
\g^2= \frac{1}{1-u^2}, 
\quad
\cot\tfrac{\pp}{2} = \frac{u}{\sqrt{\qt^2-u^2}}.
\end{equation}

\subsection{Dressing the perturbed BMN string}

To apply the dressing method to the perturbed BMN string
\begin{equation}
g_0 = g_{\rm{BMN}}  + \delta\, g_{\rm{pert}},
\end{equation}
in each step we keep terms up to first order in $\delta$. For example
\begin{equation}
g^{-1} _0= g^{-1} _{\rm{BMN}} -
 \delta\, g^{-1} _{\rm{BMN}}\, g_{\rm{pert}}\,  g^{-1} _{\rm{BMN}}.
\end{equation}
The auxiliary problem can be written as
\begin{equation}
A_0 = A_{\rm{BMN}}  + \delta\, A_{\rm{pert}},
\quad
B_0 = B_{\rm{BMN}}  + \delta\, B_{\rm{pert}},
\end{equation}
and its solution
\begin{equation}
\Psi_0(\lambda) = 
\Psi_{\rm{BMN}}(\lambda) + \delta\, \Psi_{\rm{pert}}(\lambda).
\end{equation}
Then we expand the projector
\begin{align}
\begin{split}
P_0 &= \frac{X_0}{y_0} 
= \frac{X_{\rm{BMN}}  + \delta\, X_{\rm{pert}}}{y_{\rm{BMN}}  + \delta\, y_{\rm{pert}}}
\equiv P_{\rm{BMN}} + \delta\, P_{\rm{pert}},
\end{split}
\end{align}
i.e.
\begin{equation}
P_{\rm{pert}} =  \frac{X_{\rm{pert}}}{y_{\rm{BMN}}} - \frac{y_{\rm{pert}}}{y_{\rm{BMN}}}\, P_{\rm{BMN}},
\end{equation}
and the dressing factor \eqref{fullSpec:dressedSln}
\begin{equation}
\chi_0 = \chi_{\rm{BMN}} + \delta\, \chi_{\rm{pert}}
\ : \qquad
\chi_{\rm{pert}} = \frac{\bar{\lambda}_1 - \lambda_1}{\left|\lambda_1\right|} P_{\rm{pert}}.
\end{equation}
Finally, the dressed solution is
\begin{equation}
g_1 = \chi_0\, g_0 \approx  \chi_{\rm{BMN}}\, g_{\rm{BMN}} 
	+ \delta \left( \chi_{\rm{pert}}\, g_{\rm{BMN}} + \chi_{\rm{BMN}}\, g_{\rm{pert}}\right)
\end{equation}
from which we can read off the perturbation as the first order term. Let us now apply these steps to the three perturbations we found in\footnote{
	Setting $\sin\vp = 1$ for the $\Sphere^3_{-}$ perturbations of the \AdsSSS magnon gives the $\Sphere^3$ fluctuations of the \AdsST BMN string.
} \eqref{fullSpec:SminSln1}--\eqref{fullSpec:SminSln2}.

\paragraph{Massless fluctuation.}

The massless BMN perturbation is
\begin{equation}
g_{\rm{pert}} = e^{i k x - i \w t}
	\begin{pmatrix}
		i e^{i t} 	&  0	\\
		0			&  -i e^{-i t}	\\
	\end{pmatrix},
	\qquad
	\w^2 = k^2,
\end{equation}
for which the auxiliary problem has perturbations
\begin{equation}
A_{\rm{pert}} = i (\w -k) e^{i k x - i \w t}
\begin{pmatrix}
	+1 & 0 \\
	0 & -1 \\
\end{pmatrix}, \quad
B_{\rm{pert}} = i (\w +k) e^{i k x - i \w t}
\begin{pmatrix}
	-1 & 0 \\
	0 & +1 \\
\end{pmatrix},
\end{equation}
and\vspace{-0.5em}
\begin{equation}
\Psi_{\rm{pert}}(\lambda) = 
\frac{i\ (k (1 + q \lambda) + \w \lambda )}{k (1 - (1-q)\lambda)(1 + (1+q)\lambda)}
e^{i k x - i \w t}
\begin{pmatrix}
	e^{-i Z(\lambda)} 	&  0							\\
	0							&  -e^{i Z(\lambda)}	\\
\end{pmatrix}.
\end{equation}
Then $P_{\rm{pert}}$ can be calculated in terms of
\begin{align}
\begin{split}
y_{\rm{pert}} & = - \frac{2 \qt k \sin\hp}{\w - \qt k \cos\hp} e^{i k x - i \w t} \sinh U
\\
X_{\rm{pert}} &=- \frac{i}{\w - \qt k \cos\hp}e^{i k x - i \w t} 
\begin{pmatrix}
	i \qt k \sin\hp e^{-U}	&  -(\w + q k - \qt k \cos\hp )e^{i  (t + V)}	\\
	(\w + q k - \qt k \cos\hp)e^{-i  (t + V)}	& 	-i \qt k \sin\hp e^{U}	
\end{pmatrix}
\end{split}
\end{align}
Finally, from 
\begin{equation}
\chi_{\rm{pert}}\, g_{\rm{BMN}} + \chi_{\rm{BMN}}\, g_{\rm{pert}} = 
\begin{pmatrix}
	z_1 	             &  -i z_2	    \\
	-i \bar{z}_2	&  \bar{z}_1	\\
\end{pmatrix}.
\end{equation}
we can read off the fluctuation components (after a constant rescaling)
\begin{align}
\begin{split}
z_1 & = - i e^{i k x - i \w t} e^{+i  t}
\Big(\qt k - \w \cos\hp 
\\&\qquad\qquad
- i \sin\hp\, \tanh U
\left(\w - \qt k\, \cosh (U + i \hp)\,  \sech U \right) \Big) ,
\\[0.5em]
\bar{z}_1 & =\phantom{-} i e^{i k x - i \w t} e^{-i t}
\Big(\qt k - \w \cos\tfrac{\pp}{2} 
\\&\qquad\qquad
+ i \sin\tfrac{\pp}{2}\, \tanh U
\left(\w - \qt k\, \cosh (U- i \hp)\,  \sech U \right) \Big) ,
\\[0.5em]
z_2 &= \phantom{-} i e^{i k x - i \w t}  
\sin\hp\, e^{+ i V} \sech U \left( q k -i \qt k \sin\hp\, \tanh U \right),
\\[0.5em]
\bar{z}_2 &= - i e^{i k x - i \w t}  
\sin\hp\, e^{ -i V} \sech U \left( q k +i \qt k \sin\hp\, \tanh U \right),
\end{split}
\end{align}

\paragraph{Massive fluctuation (1).}

The first massive BMN fluctuation is
\begin{equation}
g_{\rm{pert}} = e^{i k x - i \w t}
	\begin{pmatrix}
		0 	&  1	\\
		0	&  0	\\
	\end{pmatrix},
	\qquad
	\w^2 = (1 + q k)^2  + \qt^2 k^2,
\end{equation}
for which the auxiliary problem has perturbations
\begin{align}
\begin{split}
A_{\rm{pert}} &= +(\w + 1  -k) e^{i t} e^{i k x - i \w t}
\begin{pmatrix}
	0 & 1 \\
	0 & 0 \\
\end{pmatrix},
\\
B_{\rm{pert}} &= - (\w + 1  + k) e^{i t} e^{i k x - i \w t}
\begin{pmatrix}
	0 & 1 \\
	0 & 0 \\
\end{pmatrix},
\end{split}
\end{align}
and\vspace{-1.5em}
\begin{equation}
\Psi_{\rm{pert}}(\lambda) = 
\frac{e^{i t} e^{i k x - i \w t}}{1 + (1+q) \frac{\w -1 -k}{\w +1 - k} \lambda}
\begin{pmatrix}
	0 	&  e^{i Z(\lambda)}							\\
	0	&  0	\\
\end{pmatrix}.
\end{equation}
Further substituting and using the identity
\begin{equation}
\frac{1}{1 + \frac{\w -1 -k}{\w +1 - k} \sqrt{\frac{1+q}{1-q}}  e^{i \hp}}
=\frac{1}{2}\frac{\w + 1 +q k - \qt k e^{-i \hp}}{\w - \qt k \cos\hp}
\end{equation}
one gets the magnon fluctuation (rescaled by a constant)
\begin{align}
\begin{split}
z_1 & = - i e^{i k x - i \w t}e^{ -i V}  e^{+i t} 
\sin\hp\, \sech U \Big(\w + 1 + q k -  \qt k\, \cosh (U + i \hp)\, \sech U \Big) ,
\\[0.5em]
\bar{z}_1 & =- i e^{i k x - i \w t}e^{ -i V}  e^{-i t} 
\sin\hp\, \sech U \Big(\w - 1 - q k -  \qt k\, \cosh (U - i \hp)\, \sech U \Big) ,
\\[0.5em]
z_2 &= \phantom{-}i e^{i k x - i \w t} \Big( \qt k\, \sin^2\!\hp\, \sech^2\! U-2 (\qt k - \w \cos\hp ) - 2i(1+ q k) \sin\hp\, \tanh U \Big)
\\[0.5em]
\bar{z}_2 &=   \phantom{-}i e^{i k x - i \w t} e^{ -2 i V} \qt k\, \sin^2\!\hp\, \sech^2\! U.
\end{split}
\end{align}

\paragraph{Massive fluctuation (2).}

The other massive BMN fluctuation is
\begin{equation}
g_{\rm{pert}} = e^{i k x - i \w t}
	\begin{pmatrix}
		0 	&  0	\\
		1	&  0	\\
	\end{pmatrix},
	\qquad
	\w^2 = (1 - q k)^2  + \qt^2 k^2,
\end{equation}
for which the auxiliary problem has perturbations
\begin{align}
\begin{split}
A_{\rm{pert}} &= +(\w - 1  -k) e^{-i t} e^{i k x - i \w t}
\begin{pmatrix}
	0 & 0 \\
	1 & 0 \\
\end{pmatrix},
\\
B_{\rm{pert}} &= - (\w - 1  + k) e^{-i t} e^{i k x - i \w t}
\begin{pmatrix}
	0 & 0 \\
	1 & 0 \\
\end{pmatrix},
\end{split}
\end{align}
and\vspace{-1.5em}
\begin{equation}
\Psi_{\rm{pert}}(\lambda) = 
\frac{e^{- i t} e^{i k x - i \w t}}{1 + (1+q) \frac{\w +1 -k}{\w -1 - k} \lambda}
\begin{pmatrix}
	0 	  						&  0  \\
	e^{-i Z(\lambda)}	&  0	\\
\end{pmatrix}.
\end{equation}
Further substituting, using the identity
\begin{equation}
\frac{1}{1 + \frac{\w +1 -k}{\w -1 - k} \sqrt{\frac{1+q}{1-q}}  e^{i \hp}}
=\frac{1}{2}\frac{\w - 1 +q k - \qt k e^{-i \hp}}{\w - \qt k \cos\hp}
\end{equation}
and after constant rescaling, one can read off the magnon fluctuation
\begin{align}
\begin{split}
z_1 & = - i e^{i k x - i \w t}e^{i V}  e^{+i t} 
\sin\hp\, \sech U \Big(\w + 1 - q k -  \qt k\, \cosh ( U + i \hp)\, \sech U \Big) ,
\\[0.5em]
\bar{z}_1 & =- i e^{i k x - i \w t}e^{ i V}  e^{-i t} 
\sin\hp\, \sech U \Big(\w - 1 + q k -  \qt k\, \cosh ( U - i \hp)\, \sech U \Big) ,
\\[0.5em]
z_2 &=   \phantom{-}i e^{i k x - i \w t} e^{2 i V} \qt k\, \sin^2\!\hp\, \sech^2\! U,
\\[0.5em]
\bar{z}_2 &= \phantom{-}i e^{i k x - i \w t} \Big( \qt k\, \sin^2\!\hp\, \sech^2\! U-2 (\qt k - \w \cos\hp ) - 2i(1- q k) \sin\hp\, \tanh U \Big).
\end{split}
\end{align}

\section{Comparison to \texorpdfstring{$\AdS_5 \times \Sphere^5$}{AdS_5 x S^5} fluctuations}  
\label{fullSpec:AppCompareToPS}

In this appendix we compare our solutions, in the $\vp=q=0$ limit, to the 
fluctuations of the $\AdS_5 \times \Sphere^5$ giant magnon found in 
\cite{Papathanasiou:2007gd}. To harmonize notation, we need to write the
frequency and wavenumber in the boosted worldsheet basis
\begin{align}\label{fullSpec:PSbasis}
\begin{split}
e^{i k x - i \w t} &= e^{i \hat{k} \cX - i \hat{\w} \cT},
\\
k = \g (\hat{k} + u \hat{\w}) &= \csc\hp (\hat{k} + \cos\hp \hat{\w}),
\\
\w = \g (\hat{\w} + u \hat{k}) &= \csc\hp (\hat{\w} + \cos\hp \hat{k}),
\end{split}
\end{align}
where we also used the $q=0$ version of \eqref{fullSpec:pRel}.

\subsection{Bosonic fluctuations} 

Although in the $q=0$ limit the stationary magnon reduces to the HM 
giant magnon, due to obvious differences in the geometry we will only 
match a subset of our fluctuations to a subset of the ones found in
\cite{Papathanasiou:2007gd}. The magnon on $\AdS_5 \times \Sphere^5$ 
has four massive and one (unphysical) massless fluctuations on $\AdS_5$,
and, four massive and one (unphysical) massless fluctuations on 
$\Sphere^5$. Out of these, we will match both unphysical and four of 
the massive modes (two each on $\AdS_3$ and $\Sphere^3$), while our
massless modes on the $\Torus^4$ have no counterparts on 
$\AdS_5 \times \Sphere^5$.
The pure plane-wave $\AdS_3$ bosons \eqref{fullSpec:masslessAdsMode}, 
\eqref{fullSpec:massiveAdsMode} are trivially the same as the $\AdS_5$ 
bosons (2.11) of \cite{Papathanasiou:2007gd} (restricted to the 
$\AdS_3 \subset \AdS_5$ subspace), so let us focus on the $\Sphere^3$
fluctuations. Substituting \eqref{fullSpec:PSbasis}, the massless solution
\eqref{fullSpec:masslesSplusMode} becomes
\begin{align}
\begin{split}
z_1 & = - i e^{i \hat{k} \cX - i \hat{\w} \cT} e^{+i  t} \sin\hp
			\left(\hat{k} - \hat{\w} \sinh\cX \sinh(\cX + i \hp) \right),
\\[0.5em]
\bar{z}_1 & =\phantom{-} i e^{i \hat{k} \cX - i \hat{\w} \cT} e^{-i  t} \sin\hp
			\left(\hat{k} - \hat{\w} \sinh\cX \sinh(\cX - i \hp) \right),
\\[0.5em]
z_2 &= \bar{z}_2 = e^{i \hat{k} \cX - i \hat{\w} \cT} \sin\hp
				 ( \hat{k} + \cos\hp \hat{\w} )\, \sech\cY \tanh\cY,
\end{split}
\end{align}
which, up to a factor of $\sin\hp$, matches\footnote{
	Note that $\delta Z$, $\delta \vec{X}$ of \cite{Papathanasiou:2007gd} 
	are related to our notation by $z_1 = \delta Z$, 
	$z_2 = \delta X_3 + i  \delta X_4$, and we have chosen the 
	magnon-polarization vector $\vec{n}$ to point in the $X_3$ direction.
} equation (2.19) of \cite{Papathanasiou:2007gd}. In this limit the massive boson \eqref{fullSpec:bosFluc1} reduces to
\begin{align} \label{fullSpec:bosFluc1Q0}
\begin{split}
z_1 & = - e^{i \hat{k} \cX - i \hat{\w} \cT} e^{+i  t} \sin\hp\, \sech^2\!\cX
			\left( \hat{k} \sinh\cX + \hat{\w}\sinh(\cX+i\hp) + i \cosh\cX \right), 
\\[0.5em]
\bar{z}_1 & =\phantom{-} e^{i \hat{k} \cX - i \hat{\w} \cT} e^{+i  t} 
						\sin\hp\, \sech^2\!\cX
			\left( \hat{k} \sinh\cX + \hat{\w}\sinh(\cX-i\hp) + i \cosh\cX \right), 
\\[0.5em]
z_2 &= \fixedspaceR{-}{i} e^{i \hat{k} \cX - i \hat{\w} \cT} \sin\hp
			\left( ( \hat{k} + \cos\hp \hat{\w} )\, \sech^2\!\cX
			         - 2 (\hat{k} + i \tanh\cX)  \right),
\\[0.5em]
\bar{z}_2 &=  \fixedspaceR{-}{i} e^{i \hat{k} \cX - i \hat{\w} \cT} \sin\hp
			( \hat{k} + \cos\hp \hat{\w} )\, \sech^2\!\cX,
\end{split}
\end{align}
while \eqref{fullSpec:bosFluc2} becomes
\begin{align} \label{fullSpec:bosFluc2Q0}
\begin{split}
z_1 & = - e^{i \hat{k} \cX - i \hat{\w} \cT} e^{+i  t} \sin\hp\, \sech^2\!\cX
			\left( \hat{k} \sinh\cX + \hat{\w}\sinh(\cX+i\hp) + i \cosh\cX \right), 
\\[0.5em]
\bar{z}_1 & =\phantom{-} e^{i \hat{k} \cX - i \hat{\w} \cT} e^{+i  t} 
						\sin\hp\, \sech^2\!\cX
			\left( \hat{k} \sinh\cX + \hat{\w}\sinh(\cX-i\hp) + i \cosh\cX \right), 
\\[0.5em]
z_2 &=\fixedspaceR{-}{i} e^{i \hat{k} \cX - i \hat{\w} \cT} \sin\hp
			( \hat{k} + \cos\hp \hat{\w} )\, \sech^2\!\cX, 
\\[0.5em]
\bar{z}_2 &=  \fixedspaceR{-}{i} e^{i \hat{k} \cX - i \hat{\w} \cT} \sin\hp
			\left( ( \hat{k} + \cos\hp \hat{\w} )\, \sech^2\!\cX
			         - 2 (\hat{k} + i \tanh\cX)  \right).
\end{split}
\end{align}
Although the two $m=1$ bosons do not mix for $q>0$, as can be seen
from their dispersion relations $\w^2 = (1 \pm q k)^2  + \qt^2 k^2$, 
in the pure R-R limit they become degenerate, and one can take linear
combinations to match the specific solutions of \cite{Papathanasiou:2007gd}. The difference $\tfrac{1}{2}\big( \eqref{fullSpec:bosFluc2Q0} - 
\eqref{fullSpec:bosFluc1Q0}\big)$\vspace{-1em}
\begin{align}
\begin{split}
z_1 & = \bar{z}_1  = 0
\\[0.5em]
z_2 &=\phantom{-} i e^{i \hat{k} \cX - i \hat{\w} \cT} \sin\hp
			(\hat{k} + i \tanh\cX), 
\\[0.5em]
\bar{z}_2 &=  - i e^{i \hat{k} \cX - i \hat{\w} \cT} \sin\hp
			(\hat{k} + i \tanh\cX), 
\end{split}
\end{align}
reproduces the solution (2.22) of \cite{Papathanasiou:2007gd}, with $\vec{m}$ pointing along the $X_4$ direction, while the sum 
$\tfrac{1}{2}\big( \eqref{fullSpec:bosFluc1Q0} +
\eqref{fullSpec:bosFluc2Q0}\big)$
\begin{align}
\begin{split}
z_1 & = - e^{i \hat{k} \cX - i \hat{\w} \cT} e^{+i  t} \sin\hp\, \sech^2\!\cX
			\left( \hat{k} \sinh\cX + \hat{\w}\sinh(\cX+i\hp) + i \cosh\cX \right), 
\\[0.5em]
\bar{z}_1 & =\phantom{-} e^{i \hat{k} \cX - i \hat{\w} \cT} e^{+i  t} 
						\sin\hp\, \sech^2\!\cX
			\left( \hat{k} \sinh\cX + \hat{\w}\sinh(\cX-i\hp) + i \cosh\cX \right), 
\\[0.5em]
z_2 &= \bar{z}_2  = i e^{i \hat{k} \cX - i \hat{\w} \cT} \sin\hp
			\left( ( \hat{k} + \cos\hp \hat{\w} )\, \sech^2\!\cX
			         - (\hat{k} + i \tanh\cX)  \right),
\end{split}
\end{align}
matches the solution (2.20) of \cite{Papathanasiou:2007gd}, with $\vec{m} = \vec{n}$ pointing along the $X_3$ direction.

\subsection{Fermionic fluctuations} 
Since $\AdS_5 \times \Sphere^5$ is supported by 5-form fluxes, while \AdsST is supported by 3-form fluxes, the spinor structure of fermion fluctuations on the two backgrounds will be quite different, however, it is reasonable to expect similar functional forms. The kappa-fixed solutions (3.35), (3.37) in
\cite{Papathanasiou:2007gd} are of the form
\begin{align}\label{fullSpec:PSfermions}
\begin{split}
\Psi^1 &\sim \csc\tfrac{\pp}{4}\,
\sqrt{\hat{\w} + \hat{k}}\, \sech \cX \sqrt{\hat{\w} \cosh 2\cX + \hat{k}}\,
e^{i \alpha} e^{\pm i \chi}  U,
\\
\Psi^2 &\sim \sec\tfrac{\pp}{4}\,
\sqrt{\hat{\w} - \hat{k}}\, \sech \cX \sqrt{\hat{\w} \cosh 2\cX - \hat{k}}\,
e^{i \beta} e^{\pm i \tilde{\chi}} U,
\end{split}
\end{align}
where $\chi, \tilde{\chi}$ are the same as our \eqref{fullSpec:logPhasesq} 
and
\begin{align}
\begin{split}
e^{i \alpha} &= e^{i \hat{k} \cX - i \hat{\w} \cT}
					\left( \frac{1 + i \hat{\w} \sinh 2 \cX}{1 - i \hat{\w} \sinh 2 \cX}
                                \frac{1 - i \hat{k} \tanh 2 \cX}{1 + i \hat{k} \tanh 2 \cX}
                      \right)^{1/4},
\\
e^{i \beta} &= e^{i \hat{k} \cX - i \hat{\w} \cT}
					\left( \frac{1 - i \hat{\w} \sinh 2 \cX}{1 + i \hat{\w} \sinh 2 \cX}
                                \frac{1 - i \hat{k} \tanh 2 \cX}{1 + i \hat{k} \tanh 2 \cX}
                      \right)^{1/4}.
\end{split}
\end{align}
At first glance these solutions seem rather different from
\eqref{fullSpec:massiveT4fermion}, but for $\hat{\w} = \sqrt{\hat{k}^2+1}$
\begin{align}
\begin{split}
\sqrt{\hat{\w} + \hat{k}}\, \sech \cX \sqrt{\hat{\w} \cosh 2\cX + \hat{k}}\,
e^{i \alpha} & =\phantom{-} i e^{i \hat{k} \cX - i \hat{\w} \cT} 
\left( \tanh\cX - i (\hat{k} + \hat{\w}) \right),
\\
\sqrt{\hat{\w} - \hat{k}}\, \sech \cX \sqrt{\hat{\w} \cosh 2\cX - \hat{k}}\,
e^{i \beta} & =-  i e^{i \hat{k} \cX - i \hat{\w} \cT} 
\left( \tanh\cX - i (\hat{k} - \hat{\w}) \right).
\end{split}
\end{align}
\begin{equation}
\csc\tfrac{\pp}{4} =\sqrt{ \frac{2}{1-u}},
\qquad
\sec\tfrac{\pp}{4} =\sqrt{ \frac{2}{1+u}},
\end{equation}
and we can rewrite \eqref{fullSpec:PSfermions} as
\begin{align}
\begin{split}
\Psi^1 &\sim \frac{1}{\sqrt{1-u}}\, e^{i \hat{k} \cX - i \hat{\w} \cT} 
\left( \tanh\cX - i (\hat{k} + \hat{\w}) \right)e^{\pm i \chi}  U,
\\
\Psi^2 &\sim \frac{1}{\sqrt{1+u}}\, e^{i \hat{k} \cX - i \hat{\w} \cT} 
\left( \tanh\cX - i (\hat{k} - \hat{\w}) \right) e^{\pm i \tilde{\chi}} U,
\end{split}
\end{align}
in agreement with the $q=0$ limit of
\eqref{fullSpec:massiveT4fermion}, with the caveat that in 
\cite{Papathanasiou:2007gd} the spinors are swapped
$\Psi^1 \leftrightarrow \Psi^2$ compared to our notation.

\section{Coefficients in the reduced equations of motion}  \label{fullSpec:AppRedEqCoeffs}

Here we present the coefficients of the reduced equations
\eqref{fullSpec:RedEom1}, and to do so in a relatively compact form
we need to introduce the shorthands
\begin{equation}
p_{1268} = \tfrac{1}{2}(1-\lam12 \lambda_{68}),
\end{equation}
\begin{equation}
\xi = \frac{q u}{\sqrt{\qt^2-u^2}} \ ,
\end{equation}
and define
\begin{equation}
N_{ab} = \frac{i}{2}  \lam12 \Bigg(
 						 \frac{a q}{\sqrt{\qt^2 - u^2}}
 						+  \frac{Q_{b} \sqrt{1-Q_{b}^2}\ \sech^2\!\cY}
						  {1 - Q_{b}^2\ \sech^2\!\cY}
 						\Bigg),
\qquad
a,b \in \{\pm\}.
\end{equation}
With these, we have
\begin{align}
\begin{split}
C_{f_1f_1} & = + N_{-+} + i (\wt + \lam12(1 + p_{1268} \tan^2\!\vp) \xi) - i \lam12 q\, \g \zeta^{-1}\, p_{1268} \tan^2\!\vp ,
\\[0.5em]
C_{g_1g_1}  &= - N_{++}+ i (\wt + \lam12 p_{1268} \tan^2\!\vp\, \xi) - i \lam12 q\, \g \zeta^{-1}\, p_{1268} \tan^2\!\vp ,
\\[0.5em]
C_{f_2f_2} &= + N_{--} -  i (\wt + \lam12(1 + p_{1268} \tan^2\!\vp) \xi)- i \lam12 q\, \g \zeta^{-1}\, p_{1268} \tan^2\!\vp ,
\\[0.5em]
C_{g_2g_2}  &= - N_{+-} -  i (\wt + \lam12 p_{1268} \tan^2\!\vp\, \xi) - i \lam12 q\, \g \zeta^{-1}\, p_{1268} \tan^2\!\vp ,
\end{split}
\\[1em]
\begin{split}
C_{f_1f_2}   & = (1-u)\g\,
						e^{\int (-N_{-+}+ N_{--}) \d \cY}\ (1 + p_{1268} \tan^2\!\vp) (\lam12 \tanh\cY - i \xi),
\\[0.5em]
C_{g_1g_2} &= (1-u)\g\,
						e^{\int (+N_{++} - N_{+-}) \d \cY}\  p_{1268} \tan^2\!\vp (\lam12 \tanh\cY + i \xi),
\\[0.5em]
C_{f_2f_1}   &= (1+u)\g\,
						e^{\int (- N_{--}+N_{-+}) \d \cY}\ (1 + p_{1268} \tan^2\!\vp) (\lam12 \tanh\cY + i \xi),
\\[0.5em]
C_{g_2g_1} &= (1+u)\g\,
						e^{\int (+N_{+-} - N_{++}) \d \cY}\  p_{1268} \tan^2\!\vp (\lam12 \tanh\cY - i \xi),
\end{split}
\\[1em]
\begin{split}
C_{f_1g_2}   & = (1-u)\g\,
						e^{\int (-N_{-+}- N_{+-}) \d \cY}\  
						(i \lam12\, p_{1268} \tan\vp \sec\vp\, \sech \cY ),
\\[0.5em]
C_{g_1f_2} &=  (1-u)\g\,
						e^{\int (+N_{++} + N_{--}) \d \cY}\  
						(i  \lam12\, p_{1268} \tan\vp \sec\vp\, \sech \cY ),
\\[0.5em]
C_{f_2g_1}   &= (1+u)\g\,
						e^{\int (- N_{--}-N_{++}) \d \cY}\  
						(-i  \lam12\, p_{1268} \tan\vp \sec\vp\, \sech \cY ),
\\[0.5em]
C_{g_2f_1} &= (1+u)\g\,
						e^{\int (+N_{+-} + N_{-+}) \d \cY}\  
						(-i  \lam12\, p_{1268} \tan\vp \sec\vp\, \sech \cY ).
\end{split}
\end{align}
Note that $p_{1268}$ is the eigenvalue of the ansatz with respect to the 
projector $\tfrac{1}{2}(\Id + \Gh^{1268})$, and $\Delta = 0$ exactly when
$p_{1268} \tan\vp= 0$. In this case we see that the last block of coefficients 
are zero, the $P_{\pm}$ parts of the equations decouple and we have 
solutions with definite $\Gh^{012345}$ chirality. 


\bibliographystyle{nb}
\bibliography{./bibliography} 

\begin{thebibliography}{10}
\ifx\href\asklfhas\newcommand{\href}[2]{#2}\fi
\ifx\arxivref\asklfhas\newcommand{\arxivref}[2]{\href{http://arxiv.org/abs/#1}{#2}}\fi
\ifx\doiref\asklfhas\newcommand{\doiref}[2]{\href{http://dx.doi.org/#1}{#2}}\fi
\raggedright
\small
\parskip 0pt

\bibitem{Maldacena:1997re}
J.~M.~Maldacena,
\textit{``{The Large N limit of superconformal field theories and
  supergravity}''},
\textsf{\doiref{10.1023/A:1026654312961}{Int.~J.~Theor.~Phys.~38,~1113~(1999)}},
\texttt{\arxivref{hep-th/9711200}{hep-th/9711200}},
[Adv. Theor. Math. Phys.2,231(1998)].

\bibitem{Minahan:2002ve}
J.~A.~Minahan and K.~Zarembo,
\textit{``{The Bethe ansatz for N=4 superYang-Mills}''},
\textsf{\doiref{10.1088/1126-6708/2003/03/013}{JHEP~0303,~013~(2003)}},
\texttt{\arxivref{hep-th/0212208}{hep-th/0212208}}.

\bibitem{Beisert:2003tq}
N.~Beisert, C.~Kristjansen and M.~Staudacher,
\textit{``{The Dilatation operator of conformal N=4 superYang-Mills theory}''},
\textsf{\doiref{10.1016/S0550-3213(03)00406-1}{Nucl.~Phys.~B664,~131~(2003)}},
\texttt{\arxivref{hep-th/0303060}{hep-th/0303060}}.

\bibitem{Beisert:2003yb}
N.~Beisert and M.~Staudacher,
\textit{``{The N=4 SYM integrable super spin chain}''},
\textsf{\doiref{10.1016/j.nuclphysb.2003.08.015}{Nucl.~Phys.~B670,~439~(2003)}},
\texttt{\arxivref{hep-th/0307042}{hep-th/0307042}}.

\bibitem{Bena:2003wd}
I.~Bena, J.~Polchinski and R.~Roiban,
\textit{``{Hidden symmetries of the $AdS_5 \times S^5$ superstring}''},
\textsf{\doiref{10.1103/PhysRevD.69.046002}{Phys.~Rev.~D69,~046002~(2004)}},
\texttt{\arxivref{hep-th/0305116}{hep-th/0305116}}.

\bibitem{Kazakov:2004qf}
V.~A.~Kazakov, A.~Marshakov, J.~A.~Minahan and K.~Zarembo,
\textit{``{Classical/quantum integrability in AdS/CFT}''},
\textsf{\doiref{10.1088/1126-6708/2004/05/024}{JHEP~0405,~024~(2004)}},
\texttt{\arxivref{hep-th/0402207}{hep-th/0402207}}.

\bibitem{Arutyunov:2004vx}
G.~Arutyunov, S.~Frolov and M.~Staudacher,
\textit{``{Bethe ansatz for quantum strings}''},
\textsf{\doiref{10.1088/1126-6708/2004/10/016}{JHEP~0410,~016~(2004)}},
\texttt{\arxivref{hep-th/0406256}{hep-th/0406256}}.

\bibitem{Beisert:2005bm}
N.~Beisert, V.~A.~Kazakov, K.~Sakai and K.~Zarembo,
\textit{``{The Algebraic curve of classical superstrings on $AdS_5 \times
  S^5$}''},
\textsf{\doiref{10.1007/s00220-006-1529-4}{Commun.~Math.~Phys.~263,~659~(2006)}},
\texttt{\arxivref{hep-th/0502226}{hep-th/0502226}}.

\bibitem{Arutyunov:2004yx}
G.~Arutyunov and S.~Frolov,
\textit{``{Integrable Hamiltonian for classical strings on $AdS_5 \times
  S^5$}''},
\textsf{\doiref{10.1088/1126-6708/2005/02/059}{JHEP~0502,~059~(2005)}},
\texttt{\arxivref{hep-th/0411089}{hep-th/0411089}}.

\bibitem{Bethe:1931hc}
H.~Bethe,
\textit{``{On the theory of metals. 1. Eigenvalues and eigenfunctions for the
  linear atomic chain}''},
\textsf{\doiref{10.1007/BF01341708}{Z.~Phys.~71,~205~(1931)}}.

\bibitem{Faddeev:1996iy}
L.~D.~Faddeev,
\textit{``{How algebraic Bethe ansatz works for integrable model}''},
\texttt{\arxivref{hep-th/9605187}{hep-th/9605187}},
in: \textit{``{Relativistic gravitation and gravitational radiation.
  Proceedings, School of Physics, Les Houches, France, September 26-October 6,
  1995}''},
pp. 149-219p.

\bibitem{Staudacher:2004tk}
M.~Staudacher,
\textit{``{The Factorized S-matrix of CFT/AdS}''},
\textsf{\doiref{10.1088/1126-6708/2005/05/054}{JHEP~0505,~054~(2005)}},
\texttt{\arxivref{hep-th/0412188}{hep-th/0412188}}.

\bibitem{Beisert:2005tm}
N.~Beisert,
\textit{``{The SU(2|2) dynamic S-matrix}''},
\textsf{\doiref{10.4310/ATMP.2008.v12.n5.a1}{Adv.~Theor.~Math.~Phys.~12,~948~(2008)}},
\texttt{\arxivref{hep-th/0511082}{hep-th/0511082}}.

\bibitem{Beisert:2006qh}
N.~Beisert,
\textit{``{The Analytic Bethe Ansatz for a Chain with Centrally Extended
  su(2|2) Symmetry}''},
\textsf{\doiref{10.1088/1742-5468/2007/01/P01017}{J.~Stat.~Mech.~0701,~P01017~(2007)}},
\texttt{\arxivref{nlin/0610017}{nlin/0610017}}.

\bibitem{Janik:2006dc}
R.~A.~Janik,
\textit{``{The $AdS_5 \times S^5$ superstring worldsheet S-matrix and crossing
  symmetry}''},
\textsf{\doiref{10.1103/PhysRevD.73.086006}{Phys.~Rev.~D73,~086006~(2006)}},
\texttt{\arxivref{hep-th/0603038}{hep-th/0603038}}.

\bibitem{Beisert:2006ib}
N.~Beisert, R.~Hernandez and E.~Lopez,
\textit{``{A Crossing-symmetric phase for $AdS_5 \times S^5$ strings}''},
\textsf{\doiref{10.1088/1126-6708/2006/11/070}{JHEP~0611,~070~(2006)}},
\texttt{\arxivref{hep-th/0609044}{hep-th/0609044}}.

\bibitem{Beisert:2006ez}
N.~Beisert, B.~Eden and M.~Staudacher,
\textit{``{Transcendentality and Crossing}''},
\textsf{\doiref{10.1088/1742-5468/2007/01/P01021}{J.~Stat.~Mech.~0701,~P01021~(2007)}},
\texttt{\arxivref{hep-th/0610251}{hep-th/0610251}}.

\bibitem{Dorey:2007xn}
N.~Dorey, D.~M.~Hofman and J.~M.~Maldacena,
\textit{``{On the Singularities of the Magnon S-matrix}''},
\textsf{\doiref{10.1103/PhysRevD.76.025011}{Phys.~Rev.~D76,~025011~(2007)}},
\texttt{\arxivref{hep-th/0703104}{hep-th/0703104}}.

\bibitem{Volin:2009uv}
D.~Volin,
\textit{``{Minimal solution of the AdS/CFT crossing equation}''},
\textsf{\doiref{10.1088/1751-8113/42/37/372001}{J.~Phys.~A42,~372001~(2009)}},
\texttt{\arxivref{0904.4929}{arxiv:0904.4929}}.

\bibitem{Gervais:1974dc}
J.-L.~Gervais and B.~Sakita,
\textit{``{Extended Particles in Quantum Field Theories}''},
\textsf{\doiref{10.1103/PhysRevD.11.2943}{Phys.~Rev.~D11,~2943~(1975)}}.

\bibitem{Gervais:1975pa}
J.-L.~Gervais, A.~Jevicki and B.~Sakita,
\textit{``{Perturbation Expansion Around Extended Particle States in Quantum
  Field Theory. 1.}''},
\textsf{\doiref{10.1103/PhysRevD.12.1038}{Phys.~Rev.~D12,~1038~(1975)}}.

\bibitem{Gervais:1975yg}
J.-L.~Gervais, A.~Jevicki and B.~Sakita,
\textit{``{Collective Coordinate Method for Quantization of Extended
  Systems}''},
\textsf{\doiref{10.1016/0370-1573(76)90049-1}{Phys.~Rept.~23,~281~(1976)}},
in: \textit{``{A quest for symmetry: Selected works of Bunji Sakita}''},
281-293p.

\bibitem{Gervais:1976wr}
J.-L.~Gervais and A.~Jevicki,
\textit{``{Quantum Scattering of Solitons}''},
\textsf{\doiref{10.1016/0550-3213(76)90423-5}{Nucl.~Phys.~B110,~113~(1976)}}.

\bibitem{Hofman:2006xt}
D.~M.~Hofman and J.~M.~Maldacena,
\textit{``{Giant Magnons}''},
\textsf{\doiref{10.1088/0305-4470/39/41/S17}{J.~Phys.~A39,~13095~(2006)}},
\texttt{\arxivref{hep-th/0604135}{hep-th/0604135}}.

\bibitem{Metsaev:1998it}
R.~R.~Metsaev and A.~A.~Tseytlin,
\textit{``{Type IIB superstring action in $AdS_5 \times S^5$ background}''},
\textsf{\doiref{10.1016/S0550-3213(98)00570-7}{Nucl.~Phys.~B533,~109~(1998)}},
\texttt{\arxivref{hep-th/9805028}{hep-th/9805028}}.

\bibitem{Chen:2006gea}
H.-Y.~Chen, N.~Dorey and K.~Okamura,
\textit{``{Dyonic giant magnons}''},
\textsf{\doiref{10.1088/1126-6708/2006/09/024}{JHEP~0609,~024~(2006)}},
\texttt{\arxivref{hep-th/0605155}{hep-th/0605155}}.

\bibitem{Chen:2007vs}
H.-Y.~Chen, N.~Dorey and R.~F.~Lima~Matos,
\textit{``{Quantum scattering of giant magnons}''},
\textsf{\doiref{10.1088/1126-6708/2007/09/106}{JHEP~0709,~106~(2007)}},
\texttt{\arxivref{0707.0668}{arxiv:0707.0668}}.

\bibitem{Minahan:2007gf}
J.~A.~Minahan,
\textit{``{Zero modes for the giant magnon}''},
\textsf{\doiref{10.1088/1126-6708/2007/02/048}{JHEP~0702,~048~(2007)}},
\texttt{\arxivref{hep-th/0701005}{hep-th/0701005}}.

\bibitem{Papathanasiou:2007gd}
G.~Papathanasiou and M.~Spradlin,
\textit{``{Semiclassical quantization of the giant magnon}''},
\textsf{\doiref{10.1088/1126-6708/2007/06/032}{JHEP~0706,~032~(2007)}},
\texttt{\arxivref{0704.2389}{arxiv:0704.2389}}.

\bibitem{Klose:2010ki}
T.~Klose,
\textit{``{Review of AdS/CFT Integrability, Chapter IV.3: N=6 Chern-Simons and
  Strings on AdS4xCP3}''},
\textsf{\doiref{10.1007/s11005-011-0520-y}{Lett.~Math.~Phys.~99,~401~(2012)}},
\texttt{\arxivref{1012.3999}{arxiv:1012.3999}}.

\bibitem{Gauntlett:1998kc}
J.~P.~Gauntlett, R.~C.~Myers and P.~K.~Townsend,
\textit{``{Supersymmetry of rotating branes}''},
\textsf{\doiref{10.1103/PhysRevD.59.025001}{Phys.~Rev.~D59,~025001~(1998)}},
\texttt{\arxivref{hep-th/9809065}{hep-th/9809065}}.

\bibitem{Maldacena:2000hw}
J.~M.~Maldacena and H.~Ooguri,
\textit{``{Strings in $AdS_3$ and SL(2,R) WZW model 1.: The Spectrum}''},
\textsf{\doiref{10.1063/1.1377273}{J.~Math.~Phys.~42,~2929~(2001)}},
\texttt{\arxivref{hep-th/0001053}{hep-th/0001053}}.

\bibitem{Maldacena:2000kv}
J.~M.~Maldacena, H.~Ooguri and J.~Son,
\textit{``{Strings in $AdS_3$ and the SL(2,R) WZW model. Part 2. Euclidean
  black hole}''},
\textsf{\doiref{10.1063/1.1377039}{J.~Math.~Phys.~42,~2961~(2001)}},
\texttt{\arxivref{hep-th/0005183}{hep-th/0005183}}.

\bibitem{Maldacena:2001km}
J.~M.~Maldacena and H.~Ooguri,
\textit{``{Strings in $AdS_3$ and the SL(2,R) WZW model. Part 3. Correlation
  functions}''},
\textsf{\doiref{10.1103/PhysRevD.65.106006}{Phys.~Rev.~D65,~106006~(2002)}},
\texttt{\arxivref{hep-th/0111180}{hep-th/0111180}}.

\bibitem{Berkovits:1999im}
N.~Berkovits, C.~Vafa and E.~Witten,
\textit{``{Conformal field theory of AdS background with Ramond-Ramond
  flux}''},
\textsf{\doiref{10.1088/1126-6708/1999/03/018}{JHEP~9903,~018~(1999)}},
\texttt{\arxivref{hep-th/9902098}{hep-th/9902098}}.

\bibitem{Ashok:2009jw}
S.~K.~Ashok, R.~Benichou and J.~Troost,
\textit{``{Asymptotic Symmetries of String Theory on AdS(3) x S**3 with
  Ramond-Ramond Fluxes}''},
\textsf{\doiref{10.1088/1126-6708/2009/10/051}{JHEP~0910,~051~(2009)}},
\texttt{\arxivref{0907.1242}{arxiv:0907.1242}}.

\bibitem{Eberhardt:2018exh}
L.~Eberhardt and K.~Ferreira,
\textit{``{The plane-wave spectrum from the worldsheet}''},
\textsf{\doiref{10.1007/JHEP10(2018)109}{JHEP~1810,~109~(2018)}},
\texttt{\arxivref{1805.12155}{arxiv:1805.12155}}.

\bibitem{Eberhardt:2018vho}
L.~Eberhardt and K.~Ferreira,
\textit{``{Long strings and chiral primaries in the hybrid formalism}''},
\textsf{\doiref{10.1007/JHEP02(2019)098}{JHEP~1902,~098~(2019)}},
\texttt{\arxivref{1810.08621}{arxiv:1810.08621}}.

\bibitem{Babichenko:2009dk}
A.~Babichenko, B.~Stefanski,~Jr. and K.~Zarembo,
\textit{``{Integrability and the $AdS_3/CFT_2$ correspondence}''},
\textsf{\doiref{10.1007/JHEP03(2010)058}{JHEP~1003,~058~(2010)}},
\texttt{\arxivref{0912.1723}{arxiv:0912.1723}}.

\bibitem{OhlssonSax:2011ms}
O.~Ohlsson~Sax and B.~Stefanski,~Jr.,
\textit{``{Integrability, spin-chains and the $AdS_3/CFT_2$ correspondence}''},
\textsf{\doiref{10.1007/JHEP08(2011)029}{JHEP~1108,~029~(2011)}},
\texttt{\arxivref{1106.2558}{arxiv:1106.2558}}.

\bibitem{Sundin:2012gc}
P.~Sundin and L.~Wulff,
\textit{``{Classical integrability and quantum aspects of the $AdS_3\times S^3
  \times S^3 \times S^1$ superstring}''},
\textsf{\doiref{10.1007/JHEP10(2012)109}{JHEP~1210,~109~(2012)}},
\texttt{\arxivref{1207.5531}{arxiv:1207.5531}}.

\bibitem{Cagnazzo:2012se}
A.~Cagnazzo and K.~Zarembo,
\textit{``{B-field in $AdS_3/CFT_2$ Correspondence and Integrability}''},
\textsf{\doiref{10.1007/JHEP11(2012)133,
  10.1007/JHEP04(2013)003}{JHEP~1211,~133~(2012)}},
\texttt{\arxivref{1209.4049}{arxiv:1209.4049}}.

\bibitem{OhlssonSax:2018hgc}
O.~Ohlsson~Sax and B.~Stefański,
\textit{``{Closed strings and moduli in AdS$_{3}$/CFT$_{2}$}''},
\textsf{\doiref{10.1007/JHEP05(2018)101}{JHEP~1805,~101~(2018)}},
\texttt{\arxivref{1804.02023}{arxiv:1804.02023}}.

\bibitem{Sax:2014mea}
O.~Ohlsson~Sax, A.~Sfondrini and B.~Stefanski,
\textit{``{Integrability and the Conformal Field Theory of the Higgs
  branch}''},
\textsf{\doiref{10.1007/JHEP06(2015)103}{JHEP~1506,~103~(2015)}},
\texttt{\arxivref{1411.3676}{arxiv:1411.3676}}.

\bibitem{Boonstra:1998yu}
H.~J.~Boonstra, B.~Peeters and K.~Skenderis,
\textit{``{Brane intersections, anti-de Sitter space-times and dual
  superconformal theories}''},
\textsf{\doiref{10.1016/S0550-3213(98)00512-4}{Nucl.~Phys.~B533,~127~(1998)}},
\texttt{\arxivref{hep-th/9803231}{hep-th/9803231}}.

\bibitem{Gukov:2004ym}
S.~Gukov, E.~Martinec, G.~W.~Moore and A.~Strominger,
\textit{``{The Search for a holographic dual to $AdS_{3} \times S^{3} \times
  S^{3} \times S^{1}$}''},
\textsf{\doiref{10.4310/ATMP.2005.v9.n3.a3,
  10.1142/9789812775344_0035}{Adv.~Theor.~Math.~Phys.~9,~435~(2005)}},
\texttt{\arxivref{hep-th/0403090}{hep-th/0403090}}.

\bibitem{Tong:2014yna}
D.~Tong,
\textit{``{The holographic dual of $AdS_{3} \times S^{3} \times S^{3} \times
  S^{1}$}''},
\textsf{\doiref{10.1007/JHEP04(2014)193}{JHEP~1404,~193~(2014)}},
\texttt{\arxivref{1402.5135}{arxiv:1402.5135}}.

\bibitem{Eberhardt:2017pty}
L.~Eberhardt, M.~R.~Gaberdiel and W.~Li,
\textit{``{A holographic dual for string theory on $AdS_{3} \times S^{3} \times
  S^{3} \times S^{1}$}''},
\textsf{\doiref{10.1007/JHEP08(2017)111}{JHEP~1708,~111~(2017)}},
\texttt{\arxivref{1707.02705}{arxiv:1707.02705}}.

\bibitem{Eberhardt:2018ouy}
L.~Eberhardt, M.~R.~Gaberdiel and R.~Gopakumar,
\textit{``{The Worldsheet Dual of the Symmetric Product CFT}''},
\texttt{\arxivref{1812.01007}{arxiv:1812.01007}}.

\bibitem{Eberhardt:2019niq}
L.~Eberhardt and M.~R.~Gaberdiel,
\textit{``{Strings on $\text{AdS}_3 \times \text{S}^3 \times \text{S}^3 \times
  \text{S}^1$}''},
\textsf{\doiref{10.1007/JHEP06(2019)035}{JHEP~1906,~035~(2019)}},
\texttt{\arxivref{1904.01585}{arxiv:1904.01585}}.

\bibitem{Borsato:2013qpa}
R.~Borsato, O.~Ohlsson~Sax, A.~Sfondrini, B.~Stefański and A.~Torrielli,
\textit{``{The all-loop integrable spin-chain for strings on $AdS_3 \times S^3
  \times T^4$: the massive sector}''},
\textsf{\doiref{10.1007/JHEP08(2013)043}{JHEP~1308,~043~(2013)}},
\texttt{\arxivref{1303.5995}{arxiv:1303.5995}}.

\bibitem{Hoare:2013ida}
B.~Hoare and A.~A.~Tseytlin,
\textit{``{Massive S-matrix of $AdS_3 \times S^3 \times T^4$ superstring theory
  with mixed 3-form flux}''},
\textsf{\doiref{10.1016/j.nuclphysb.2013.04.024}{Nucl.~Phys.~B873,~395~(2013)}},
\texttt{\arxivref{1304.4099}{arxiv:1304.4099}}.

\bibitem{Lloyd:2014bsa}
T.~Lloyd, O.~Ohlsson~Sax, A.~Sfondrini and B.~Stefański,~Jr.,
\textit{``{The complete worldsheet S-matrix of superstrings on $AdS_3 \times
  S^3 \times T^4$ with mixed three-form flux}''},
\textsf{\doiref{10.1016/j.nuclphysb.2014.12.019}{Nucl.~Phys.~B891,~570~(2015)}},
\texttt{\arxivref{1410.0866}{arxiv:1410.0866}}.

\bibitem{Hoare:2013lja}
B.~Hoare, A.~Stepanchuk and A.~A.~Tseytlin,
\textit{``{Giant magnon solution and dispersion relation in string theory in
  $AdS_3 \times S^3 \times T^4$ with mixed flux}''},
\textsf{\doiref{10.1016/j.nuclphysb.2013.12.011}{Nucl.~Phys.~B879,~318~(2014)}},
\texttt{\arxivref{1311.1794}{arxiv:1311.1794}}.

\bibitem{Stepanchuk:2014kza}
A.~Stepanchuk,
\textit{``{String theory in $AdS_3 \times S^3 \times T^4$ with mixed flux:
  semiclassical and 1-loop phase in the S-matrix}''},
\textsf{\doiref{10.1088/1751-8113/48/19/195401}{J.~Phys.~A48,~195401~(2015)}},
\texttt{\arxivref{1412.4764}{arxiv:1412.4764}}.

\bibitem{Babichenko:2014yaa}
A.~Babichenko, A.~Dekel and O.~Ohlsson~Sax,
\textit{``{Finite-gap equations for strings on $AdS_3 \times S^3 \times T^4$
  with mixed 3-form flux}''},
\textsf{\doiref{10.1007/JHEP11(2014)122}{JHEP~1411,~122~(2014)}},
\texttt{\arxivref{1405.6087}{arxiv:1405.6087}}.

\bibitem{Abbott:2012dd}
M.~C.~Abbott,
\textit{``{Comment on Strings in $AdS_3\times S^3 \times S^3 \times S^1$ at One
  Loop}''},
\textsf{\doiref{10.1007/JHEP02(2013)102}{JHEP~1302,~102~(2013)}},
\texttt{\arxivref{1211.5587}{arxiv:1211.5587}}.

\bibitem{Sundin:2014ema}
P.~Sundin and L.~Wulff,
\textit{``{One- and two-loop checks for the $AdS_3$ x $S^3$ x $T^4$ superstring
  with mixed flux}''},
\textsf{\doiref{10.1088/1751-8113/48/10/105402}{J.~Phys.~A~48,~105402~(2015)}},
\texttt{\arxivref{1411.4662}{arxiv:1411.4662}}.

\bibitem{Borsato:2014hja}
R.~Borsato, O.~Ohlsson~Sax, A.~Sfondrini and B.~Stefanski,
\textit{``{The complete $AdS_3 \times S^3 \times T^4$ worldsheet S-matrix}''},
\textsf{\doiref{10.1007/JHEP10(2014)066}{JHEP~1410,~66~(2014)}},
\texttt{\arxivref{1406.0453}{arxiv:1406.0453}}.

\bibitem{Borsato:2014exa}
R.~Borsato, O.~Ohlsson~Sax, A.~Sfondrini and B.~Stefanski,
\textit{``{Towards the All-Loop Worldsheet S-Matrix for $AdS_3\times S^3\times
  T^4$}''},
\textsf{\doiref{10.1103/PhysRevLett.113.131601}{Phys.~Rev.~Lett.~113,~131601~(2014)}},
\texttt{\arxivref{1403.4543}{arxiv:1403.4543}}.

\bibitem{Borsato:2012ud}
R.~Borsato, O.~Ohlsson~Sax and A.~Sfondrini,
\textit{``{A dynamic $su(1|1)^2$ S-matrix for $AdS_3/CFT_2$}''},
\textsf{\doiref{10.1007/JHEP04(2013)113}{JHEP~1304,~113~(2013)}},
\texttt{\arxivref{1211.5119}{arxiv:1211.5119}}.

\bibitem{Borsato:2015mma}
R.~Borsato, O.~Ohlsson~Sax, A.~Sfondrini and B.~Stefański,
\textit{``{The $AdS_3\times S^3 \times S^3 \times S^1$ worldsheet S-matrix}''},
\textsf{\doiref{10.1088/1751-8113/48/41/415401}{J.~Phys.~A48,~415401~(2015)}},
\texttt{\arxivref{1506.00218}{arxiv:1506.00218}}.

\bibitem{Varga:2019hqh}
A.~Varga,
\textit{``{Fermion zero modes for the mixed-flux AdS$_{3}$ giant magnon}''},
\textsf{\doiref{10.1007/JHEP02(2019)135}{JHEP~1902,~135~(2019)}},
\texttt{\arxivref{1901.00530}{arxiv:1901.00530}}.

\bibitem{Dashen:1975hd}
R.~F.~Dashen, B.~Hasslacher and A.~Neveu,
\textit{``{The Particle Spectrum in Model Field Theories from Semiclassical
  Functional Integral Techniques}''},
\textsf{\doiref{10.1103/PhysRevD.11.3424}{Phys.~Rev.~D11,~3424~(1975)}}.

\bibitem{Zakharov:1973pp}
V.~E.~Zakharov and A.~V.~Mikhailov,
\textit{``{Relativistically Invariant Two-Dimensional Models in Field Theory
  Integrable by the Inverse Problem Technique. (In Russian)}''},
\textsf{Sov.~Phys.~JETP~47,~1017~(1978)},
[Zh. Eksp. Teor. Fiz.74,1953(1978)].

\bibitem{Harnad:1983we}
J.~P.~Harnad, Y.~Saint~Aubin and S.~Shnider,
\textit{``{Backlund Transformations for Nonlinear $\sigma$ Models With Values
  in Riemannian Symmetric Spaces}''},
\textsf{\doiref{10.1007/BF01210726}{Commun.~Math.~Phys.~92,~329~(1984)}}.

\bibitem{Spradlin:2006wk}
M.~Spradlin and A.~Volovich,
\textit{``{Dressing the Giant Magnon}''},
\textsf{\doiref{10.1088/1126-6708/2006/10/012}{JHEP~0610,~012~(2006)}},
\texttt{\arxivref{hep-th/0607009}{hep-th/0607009}}.

\bibitem{Frolov:2002av}
S.~Frolov and A.~A.~Tseytlin,
\textit{``{Semiclassical quantization of rotating superstring in AdS(5) x
  S**5}''},
\textsf{\doiref{10.1088/1126-6708/2002/06/007}{JHEP~0206,~007~(2002)}},
\texttt{\arxivref{hep-th/0204226}{hep-th/0204226}}.

\bibitem{Park:2005ji}
I.~Y.~Park, A.~Tirziu and A.~A.~Tseytlin,
\textit{``{Spinning strings in AdS(5) x S**5: One-loop correction to energy in
  SL(2) sector}''},
\textsf{\doiref{10.1088/1126-6708/2005/03/013}{JHEP~0503,~013~(2005)}},
\texttt{\arxivref{hep-th/0501203}{hep-th/0501203}}.

\bibitem{Cvetic:1999zs}
M.~Cvetic, H.~Lu, C.~N.~Pope and K.~S.~Stelle,
\textit{``{T duality in the Green-Schwarz formalism, and the massless / massive
  IIA duality map}''},
\textsf{\doiref{10.1016/S0550-3213(99)00740-3}{Nucl.~Phys.~B573,~149~(2000)}},
\texttt{\arxivref{hep-th/9907202}{hep-th/9907202}}.

\bibitem{Jevicki:1979nr}
A.~Jevicki,
\textit{``{Classical And Quantum Dynamics Of Two-dimensional Nonlinear Field
  Theories: A Review }''}.

\bibitem{Borsato:2016kbm}
R.~Borsato, O.~Ohlsson~Sax, A.~Sfondrini and B.~Stefański,
\textit{``{On the spectrum of $AdS_3 \times S^3 \times T^4$ strings with
  Ramond–Ramond flux}''},
\textsf{\doiref{10.1088/1751-8113/49/41/41LT03}{J.~Phys.~A49,~41LT03~(2016)}},
\texttt{\arxivref{1605.00518}{arxiv:1605.00518}}.

\bibitem{Baggio:2017kza}
M.~Baggio, O.~Ohlsson~Sax, A.~Sfondrini, B.~Stefański and A.~Torrielli,
\textit{``{Protected string spectrum in AdS$_{3}$/CFT$_{2}$ from worldsheet
  integrability}''},
\textsf{\doiref{10.1007/JHEP04(2017)091}{JHEP~1704,~091~(2017)}},
\texttt{\arxivref{1701.03501}{arxiv:1701.03501}}.

\bibitem{deBoer:1998kjm}
J.~de~Boer,
\textit{``{Six-dimensional supergravity on S**3 x AdS(3) and 2-D conformal
  field theory}''},
\textsf{\doiref{10.1016/S0550-3213(99)00160-1}{Nucl.~Phys.~B548,~139~(1999)}},
\texttt{\arxivref{hep-th/9806104}{hep-th/9806104}}.

\bibitem{Eberhardt:2017fsi}
L.~Eberhardt, M.~R.~Gaberdiel, R.~Gopakumar and W.~Li,
\textit{``{BPS spectrum on AdS$_3\times $S$^3 \times $S$^3 \times $S$^1$}''},
\textsf{\doiref{10.1007/JHEP03(2017)124}{JHEP~1703,~124~(2017)}},
\texttt{\arxivref{1701.03552}{arxiv:1701.03552}}.

\bibitem{Baggio:2018gct}
M.~Baggio and A.~Sfondrini,
\textit{``{Strings on NS-NS Backgrounds as Integrable Deformations}''},
\textsf{\doiref{10.1103/PhysRevD.98.021902}{Phys.~Rev.~D~98,~021902~(2018)}},
\texttt{\arxivref{1804.01998}{arxiv:1804.01998}}.

\bibitem{Dei:2018mfl}
A.~Dei and A.~Sfondrini,
\textit{``{Integrable spin chain for stringy Wess-Zumino-Witten models}''},
\textsf{\doiref{10.1007/JHEP07(2018)109}{JHEP~1807,~109~(2018)}},
\texttt{\arxivref{1806.00422}{arxiv:1806.00422}}.

\bibitem{Dei:2018yth}
A.~Dei, M.~R.~Gaberdiel and A.~Sfondrini,
\textit{``{The plane-wave limit of $\text{AdS}_3 \times \text{S}^3 \times
  \text{S}^3 \times \text{S}^1$}''},
\textsf{\doiref{10.1007/JHEP08(2018)097}{JHEP~1808,~097~(2018)}},
\texttt{\arxivref{1805.09154}{arxiv:1805.09154}}.

\bibitem{Dei:2018jyj}
A.~Dei and A.~Sfondrini,
\textit{``{Integrable S matrix, mirror TBA and spectrum for the stringy
  $\text{AdS}_3 \times \text{S}^3 \times \text{S}^3 \times \text{S}^1$ WZW
  model}''},
\textsf{\doiref{10.1007/JHEP02(2019)072}{JHEP~1902,~072~(2019)}},
\texttt{\arxivref{1812.08195}{arxiv:1812.08195}}.

\bibitem{Giribet:2018ada}
G.~Giribet, C.~Hull, M.~Kleban, M.~Porrati and E.~Rabinovici,
\textit{``{Superstrings on AdS$_{3}$ at $k =$ 1}''},
\textsf{\doiref{10.1007/JHEP08(2018)204}{JHEP~1808,~204~(2018)}},
\texttt{\arxivref{1803.04420}{arxiv:1803.04420}}.

\bibitem{Gaberdiel:2018rqv}
M.~R.~Gaberdiel and R.~Gopakumar,
\textit{``{Tensionless string spectra on AdS$_{3}$}''},
\textsf{\doiref{10.1007/JHEP05(2018)085}{JHEP~1805,~085~(2018)}},
\texttt{\arxivref{1803.04423}{arxiv:1803.04423}}.

\bibitem{Eberhardt:2020akk}
L.~Eberhardt,
\textit{``{AdS$_{3}$/CFT$_{2}$ at higher genus}''},
\textsf{\doiref{10.1007/JHEP05(2020)150}{JHEP~2005,~150~(2020)}},
\texttt{\arxivref{2002.11729}{arxiv:2002.11729}}.

\bibitem{Nieto:2018jzi}
J.~M.~Nieto and R.~Ruiz,
\textit{``{One-loop quantization of rigid spinning strings in $AdS_3 \times S^3
  \times T^4$ with mixed flux}''},
\textsf{\doiref{10.1007/JHEP07(2018)141}{JHEP~1807,~141~(2018)}},
\texttt{\arxivref{1804.10477}{arxiv:1804.10477}}.

\bibitem{Sax:2012jv}
O.~Ohlsson~Sax, j.~Stefanski,~Bogdan and A.~Torrielli,
\textit{``{On the massless modes of the $AdS_3/CFT_2$ integrable systems}''},
\textsf{\doiref{10.1007/JHEP03(2013)109}{JHEP~1303,~109~(2013)}},
\texttt{\arxivref{1211.1952}{arxiv:1211.1952}}.

\bibitem{Abbott:2015pps}
M.~C.~Abbott and I.~Aniceto,
\textit{``{Massless Lüscher terms and the limitations of the AdS$_3$
  asymptotic Bethe ansatz}''},
\textsf{\doiref{10.1103/PhysRevD.93.106006}{Phys.~Rev.~D93,~106006~(2016)}},
\texttt{\arxivref{1512.08761}{arxiv:1512.08761}}.

\bibitem{Bombardelli:2018jkj}
D.~Bombardelli, B.~Stefański and A.~Torrielli,
\textit{``{The low-energy limit of AdS$_{3}$/CFT$_{2}$ and its TBA}''},
\textsf{\doiref{10.1007/JHEP10(2018)177}{JHEP~1810,~177~(2018)}},
\texttt{\arxivref{1807.07775}{arxiv:1807.07775}}.

\bibitem{Fontanella:2019ury}
A.~Fontanella, O.~Ohlsson~Sax, B.~Stefa\'nski, Jr. and A.~Torrielli,
\textit{``{The effectiveness of relativistic invariance in AdS$_{3}$}''},
\textsf{\doiref{10.1007/JHEP07(2019)105}{JHEP~1907,~105~(2019)}},
\texttt{\arxivref{1905.00757}{arxiv:1905.00757}}.

\bibitem{Abbott:2020jaa}
M.~C.~Abbott and I.~Aniceto,
\textit{``{Integrable Field Theories with an Interacting Massless Sector}''},
\texttt{\arxivref{2002.12060}{arxiv:2002.12060}}.

\end{thebibliography}

\end{document}